\documentclass[lettersize,journal]{IEEEtran}
\def \sn {\textsc{SParse}}
\usepackage{algorithm}
\usepackage{algorithmic}
\usepackage{amsfonts}
\usepackage{amsmath}
\usepackage{textcomp}
\usepackage{array}
\usepackage{xcolor}
\usepackage{epsfig}
\PassOptionsToPackage{hyphens}{url}
\usepackage{svg}
\usepackage{caption}
\usepackage{color}

\usepackage{dingbat}
\usepackage{stfloats}
\usepackage{tabularx}
\usepackage{mdwmath}
\usepackage{multirow}
\usepackage{enumerate}

\usepackage{amssymb}
\usepackage{breakurl}
\usepackage{graphicx}
\usepackage{wasysym}
\usepackage{epstopdf}
\usepackage{subfigure}
\usepackage{mathrsfs}

\usepackage{multirow}
\usepackage{CJKutf8}

\usepackage{url}

\usepackage{threeparttable}
\def\BibTeX{{\rm B\kern-.05em{\sc i\kern-.025em b}\kern-.08em
    T\kern-.1667em\lower.7ex\hbox{E}\kern-.125emX}}
\usepackage{balance}

\begin{document}
\title{SPARSE: Semantic Tracking and Path Analysis for Attack Investigation in Real-time}

\author{Jie Ying,
        Tiantian Zhu*,
        Wenrui Cheng,
        Qixuan Yuan,
        Mingjun Ma,
        Chunlin Xiong,
        Tieming Chen,
        Mingqi Lv,
        and Yan Chen, IEEE Fellow
		
}


\maketitle

\begin{abstract}
As the complexity and destructiveness of Advanced Persistent Threat (APT) increase, there is a growing tendency to identify a series of actions undertaken to achieve the attacker's target, called \textit{attack investigation}. 
Currently, analysts construct the provenance graph to perform causality analysis on Point-Of-Interest (POI) event for capturing critical events (related to the attack).
However, due to the vast size of the provenance graph and the rarity of critical events, existing attack investigation methods suffer from problems of high false positives, high overhead, and high latency.
\par
To this end, we propose~\sn{}, an efficient and real-time system for constructing critical component graphs (i.e., consisting of critical events) from streaming logs. Our key observation is 1) Critical events exist in a suspicious semantic graph (SSG) composed of interaction flows between suspicious entities, and 2) Information flows that accomplish attacker's goal exist in the form of paths. Therefore, \sn{} uses a two-stage framework to implement attack investigation (i.e., constructing the SSG and performing path-level contextual analysis).
First, \sn{} operates in a state-based mode where events are consumed as streams, allowing easy access to the SSG related to the POI event through semantic transfer rule and storage strategy. Then, \sn{} identifies all suspicious flow paths (SFPs) related to the POI event from the SSG, quantifies the influence of each path to filter irrelevant events. Our evaluation on a real large-scale attack dataset shows that \sn{} can generate a critical component graph ($\sim$ 113 edges) in 1.6 seconds, which is 2014 $\times$ smaller than the backtracking graph ($\sim$ 227,589 edges). \sn{} is 25 $\times$ more effective than other state-of-the-art techniques in filtering irrelevant edges.

\end{abstract}


\section{Introduction}\label{sec:1_intro}
As the Internet has developed over time, APT attacks have grown more sophisticated and destructive. 
APT attacks target mainly large corporations such as Twitter~\cite{twitter-200M}, resulting in significant financial losses and reputational damage. In addition, APT attacks are executed in multiple stages, which include initial access, persistence, lateral movement, collection, and exfiltration~\cite{mitre-attck}. 
\par
While an intrusion may be noticed at any stage, detection only uncovers isolated traces of the attack. As a result, analysts must undertake \textbf{causality analysis} to capture the bigger picture and obtain a sound understanding of the detected attack point. Achieving a secure system recovery after a cyber attack requires certain key steps. First, analysts must determine how the adversary infiltrated the system. Once the point of entry is identified, then analysts need to assess the obvious and hidden damage done to the system, such as installed payload, modified files, and exfiltrated information. In short, analysts need to identify the sequence of \textit{critical events} leading up to the POI event and reconstruct the \textit{critical component} (subgraph consisting of critical events), which is also called \textbf{attack investigation}.
\par
With the improvement of kernel-level monitoring frameworks~\cite{laus, etw, gehani2012spade}, more and more causality analysis systems depend on a provenance graph consisting of entities (e.g., files, processes, and sockets) and inter-entity interactions (e.g., processes reading and writing files). However, the auditing framework is known to generate a large number of logs, up to several gigabytes per day on a single machine~\cite{ma2018kernel, bates2015trustworthy}, resulting in a massive graph with billions of edges. This leads to critical events that cause the attack to be drowned out by irrelevant events of normal behavior. Also, a provenance graph is a coarse-grained data format that cannot directly determine the specific dependencies between all relevant events of an entity (e.g., a process has multiple read-in events and write-out events)~\cite{inam2022sok}.  The dependency explosion problem~\cite{lee2013high, tang2018nodemerge, xu2016high} caused by these conditions leads to the poor performance of existing causality analysis systems.

\begin{table*}[]
\centering
\caption{\centering Comparison table of related work on attack investigation performance. Column 5 (Storage of Historical Data) indicates whether to store the raw audit logs. The solidness of the marked circle reflects the degree: High (\CIRCLE), Medium (\RIGHTcircle), Low (\Circle).}
\label{tab:related work}
\resizebox{\linewidth}{!}{%
\begin{tabular}{c|c|c|c|c|c|c}
\hline
\textbf{Technique}                                                                 & \textbf{System} & \textbf{\begin{tabular}[c]{@{}c@{}}False Positive\\ Rate\end{tabular}} & \textbf{\begin{tabular}[c]{@{}c@{}}False Negative\\ Rate\end{tabular}} & \textbf{\begin{tabular}[c]{@{}c@{}}Storage of \\ Historical Data\end{tabular}} & \textbf{\begin{tabular}[c]{@{}c@{}}Memory \\ Overhead\end{tabular}} & \textbf{\begin{tabular}[c]{@{}c@{}}Time\\ Overhead\end{tabular}} \\ \hline
\multirow{3}{*}{\begin{tabular}[c]{@{}c@{}}Label \\ Propagation-based\end{tabular}} & HOLMES~\cite{milajerdi2019holmes}         & \RIGHTcircle                                                           & \CIRCLE                                                           & \Circle                                                                       & \CIRCLE                                                             & \Circle                                                          \\
                                                                                    & RapSheet~\cite{hassan2020rapsheet}        & \RIGHTcircle                                                           & \CIRCLE                                                           & \CIRCLE                                                                       & \CIRCLE                                                             & \CIRCLE                                                          \\
                                                                                    & APTSHIELD~\cite{zhu2023aptshield}       & \RIGHTcircle                                                           & \CIRCLE                                                           & \Circle                                                                        & \Circle                                                             & \Circle                                                          \\ \hline
\multirow{4}{*}{\begin{tabular}[c]{@{}c@{}}Anomaly \\ Score-based\end{tabular}}     & NODOZE~\cite{hassan2019nodoze}          & \CIRCLE                                                                & \Circle                                                                & \CIRCLE                                                                        & \CIRCLE                                                             & \CIRCLE                                                          \\
                                                                                    & Swift~\cite{hassan2020we}           & \CIRCLE                                                                & \Circle                                                                & \CIRCLE                                                                        & \CIRCLE                                                             & \CIRCLE                                                          \\
                                                                                    & PRIOTRACKER~\cite{liu2018priortracker}    & \CIRCLE                                                                & \Circle                                                                & \CIRCLE                                                                        & \CIRCLE                                                             & \CIRCLE                                                          \\
                                                                                    & DEPIMPECT~\cite{fang2022depimpact}       & \RIGHTcircle                                                           & \Circle                                                                & \CIRCLE                                                                        & \CIRCLE                                                             & \CIRCLE                                                          \\ \hline
\multirow{3}{*}{\begin{tabular}[c]{@{}c@{}}Machine \\ Learning-based\end{tabular}}  & ATLAS~\cite{alsaheel2021atlas}           & \Circle                                                                & \RIGHTcircle                                                           & \CIRCLE                                                                        & \RIGHTcircle                                                        & \RIGHTcircle                                                     \\
                                                                                    & DEPCOMM~\cite{xu2022depcomm}         & \RIGHTcircle                                                                & \CIRCLE                                                                & \CIRCLE                                                                        & \RIGHTcircle                                                        & \RIGHTcircle                                                     \\
                                                                                    & WASTON~\cite{zeng2021watson}          & \RIGHTcircle                                                           & \CIRCLE                                                                & \CIRCLE                                                                        & \RIGHTcircle                                                        & \RIGHTcircle                                                     \\ \hline
                                                /                     & \sn{}           & \Circle                                                                & \Circle                                                                & \Circle                                                                        & \Circle                                                             & \Circle                                                          \\  \hline 
                                           
\end{tabular}%
}
\end{table*}

\par
Methodologically, causality analysis can be classified into three categories: label propagation-based, anomaly score-based, and machine learning-based. Specifically, the label propagation-based approach~\cite{milajerdi2019holmes, hassan2020tactical, xiong2020conan, zhu2023aptshield} sets entity labels and transformation rules through heuristic rules but suffers from a reliance on heavy manual effort and the incapacity to address zero-day vulnerabilities. The anomaly score-based approach~\cite{fang2022depimpact, hassan2019nodoze, liu2018priortracker, hassan2020we} quantifies the suspiciousness of dependency between entities, but faces challenges such as relying on historical statistics and the inability to adapt to complex enterprise production environments. The machine learning-based approach~\cite{alsaheel2021atlas, xu2022depcomm, zeng2021watson} employs neural networks to learn from attack samples but is hindered by insufficient sample size, poor generalization capability, and high computational overhead. These issues, as shown in Table~\ref{tab:related work}, make it challenging for analysts to conduct attack investigation within the optimal time (10 minutes)~\cite{lateral-movement} while handling massive alerts.
In summary, a general, efficient, and cost-effective causality analysis system needs to meet the following three requirements: \textbf{1) Reduced False Positives} to address dependency explosion, \textbf{2) Affordable Overhead} to reduce the cost of attack investigation, and \textbf{3) Minimal Latency} to prevent further losses caused by subsequent attacks.

\par
\textbf{Key Insight.} To meet the above requirements, after researching hundreds of APT attack descriptions~\cite{apt-notes} and analyzing numerous related dependency graphs~\cite{kwon2018mci, gao2018aiql, ma2016protracer, liu2018priortracker, fang2022depimpact, alsaheel2021atlas}, we have the following two key insights. Firstly, \textit{critical events must exist in the suspicious semantic graph (SSG) formed by interactions between suspicious entities}. Specifically, we construct the SSG consisting of suspicious entities (e.g., a process visiting an unknown website) and suspicious events (data and control flows initiated by the suspicious entities). We believe that the SSG contains all critical events (i.e., critical events are a subset of suspicious events) and is much smaller than the subgraph obtained by backtracking~\cite{king2003backtracking} from the POI event, as shown in Figure~\ref{fig:dependency graph}. Secondly, \textit{information flows that accomplish goals exist in the form of paths}. In other words, we opine that evaluating whether to filter an event cannot be done in isolation, but rather calls for a comprehensive assessment of flow paths consisting of multiple events. Consequently, we construct all \textit{suspicious flow paths (SFPs)} related to the POI event from SSG and quantify the degree of influence based on the properties of the POI event and path structural characteristics to weed out irrelevant events. In summary, we achieve the attack investigation through a two-stage step (i.e., SSG construction and path-level contextual analysis).
\par
In summary, this paper proposes \sn{}\footnote{\sn{} short for \underline{S}emantic tracking and \underline{P}ath \underline{A}nalysis fo\underline{R} attack inve\underline{S}tigation in real-tim\underline{E}} and makes the following contributions:
\begin{itemize}

\item {We propose a state-based framework that contains \textit{suspicious semantic transfer rule} and \textit{suspicious event storage strategy}. The framework consumes events as streams in low overhead without recording historical data. In addition, the framework can output suspicious semantic graph related to the POI event in real time. The graph consists of all suspicious data flows and control flows that lead to the POI event, and thus contains all attack-related critical events. It is phase \uppercase\expandafter{\romannumeral1} of \sn{} for filtering semantic-irrelevant events.
}

\item {We propose a path-level contextual analysis mechanism that incorporates \textit{suspicious flow path extraction and scoring}. It utilizes an optimized BFS algorithm to extract all suspicious flow paths (SFPs) from the SSG. Then the mechanism combines the properties of the POI event and characteristics of the path structure to quantify the impact of each SFP on the POI event. Finally, it filters all events that only exist in SFPs with low scores. It is phase \uppercase\expandafter{\romannumeral2} of \sn{} for filtering impact-irrelevant events.}

\item {We implemented \sn{} and evaluated all its components in detail on a large-scale dataset with more than 150 million logs. Specifically, the dataset contains 10 simulated attacks~\cite{fang2022depimpact} ($ \sim $ 100 million logs) and 5 attacks from the DARPA TC program~\cite{darpa-tc, darpa3_dataset} ($ \sim $ 50 million logs). Experimental results show that \sn{} can generate the critical component graph ($\sim$ 113 edges) in 1.6s, which is 2014 $\times$ smaller than the dependency graph ($\sim$ 227,589 edges). The critical component graph (FP = 99) generated by \sn{} is 25 $\times$ more effective than other state-of-the-art causality analysis techniques (FP = 2,473) in filtering irrelevant edges while preserving the attack sequences. In addition, \sn{} can run for a long time while processing streaming logs with a low memory overhead (30MB).}

\end{itemize}

\section{Background and Motivation}
\label{sec:2_background}
\subsection{Dependency Graph}
\label{sec:2.1_dataformat}
Recent literature has leveraged the concept of \textit{data provenance}, i.e., instead of manually piecing together individual evidence from raw logs, provenance-based systems can construct \textit{dependency graphs} that explain the relationships between each event, simplifying the attack investigation. Specifically, a dependency graph $G(E,\ V)$ is a heterogeneous graph consisting of nodes $V$ representing system entities and edges $E$ representing inter-entity events. The attributes of entities and events are carefully selected from raw audit logs, which are lean and critical. For entities, we choose processes ($\langle ProcessName,\ Process ID \rangle$) , files ($\langle FileName \rangle$), and sockets ($\langle IP\ :\ Port \rangle$). For events, we made selections as shown in Table~\ref{tab:attribute_Event}. For any edge $e \in E$, there is $e = (u, v, t)$, where $u$ represents the subject, $v$ represents the object, and $t$ represents the timestamp of the event. For the two edges in the dependency graph, $e1 = (u_{1}, v_{1}, t_{1})$, $e2 = (u_{2}, v_{2}, t_{2})$, we consider that there is a dependence (causality) between $e_{1}$ and $e_{2}$ if $v_{1} = u_{2}$ and $t_{1}  < t_{2}$.


\begin{table}[]
\centering
\caption{\textbf{Attributes of system events}}
\label{tab:attribute_Event}
\resizebox{0.45\textwidth}{!}{%
\begin{tabular}{c|c|c}
\hline
\textbf{Event} & \textbf{Operations}                                                              & \textbf{Attributes}                                                                                            \\ \hline
Process Event  & \begin{tabular}[c]{@{}c@{}}execve, clone\end{tabular}                   & \multirow{3}{*}{\begin{tabular}[c]{@{}c@{}}Time Stamp\\ Subject Name\\ Object Name\\ Data Amount\end{tabular}} \\ \cline{1-2}
File Event     & \begin{tabular}[c]{@{}c@{}}write, read, \\ readv, writev\end{tabular}     &                                                                                                                \\ \cline{1-2}
Network Event  & \begin{tabular}[c]{@{}c@{}} sendto, recvfrom\end{tabular} &                                                                                                                \\ \hline
\end{tabular}%
}
\end{table}

\begin{figure*}[h!t]
\centering
\includegraphics[width=\linewidth]{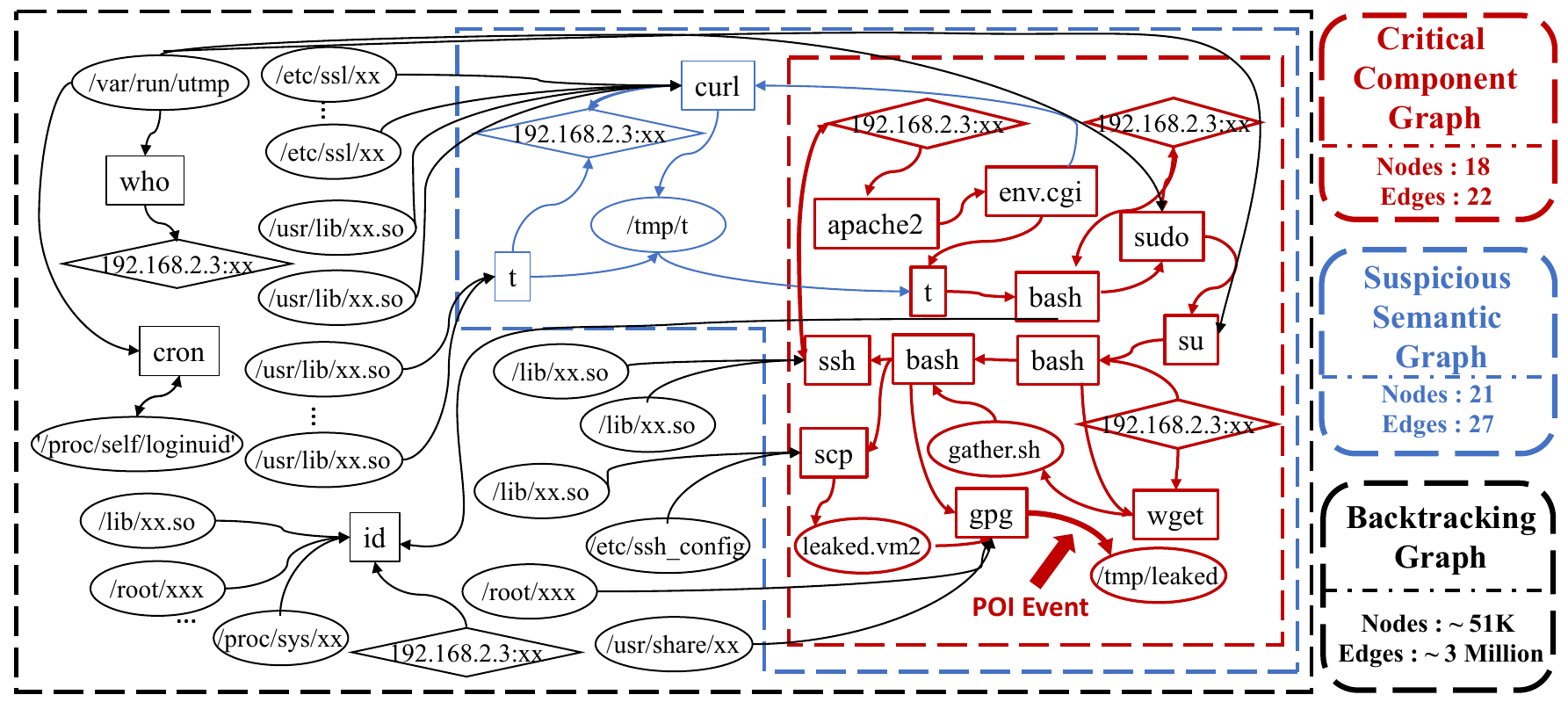}
\caption{Partial dependency graph of one attack case \textit{Dataleak}. 
The black dashed box indicates the backtracking graph ($\sim$ 200,000 edges) constructed from the POI event via backward propagation. The blue dashed box indicates the suspicious semantic graph ($\sim$ 27 edges) constructed by \sn{}. The red dashed line box indicates the critical component graph ($\sim$ 22 edges) exported by \sn{}.}
\label{fig:dependency graph}
\end{figure*}

\subsection{Attack Investigation}
\par
The goal of attack investigation using dependency graphs~\cite{milajerdi2019holmes, hassan2019nodoze, hassan2020rapsheet, hassan2020we, xu2022depcomm, fang2022depimpact, zhu2023aptshield, xiong2020conan, zhu2021gs-ss} is to identify all \textit{critical events} and \textit{critical components} related to the POI event. A critical component is a subgraph of the dependency graph that retains internal information critical to the attack investigation and eliminates irrelevant system activity. Typically this analysis includes tracing the flow of data through the graph to identify potentially relevant events, and examining the properties of nodes and edges to identify signs of compromise. The goal of an attack investigation is to determine both the source and scope of the attack, ascertain the extent of damage or disruption, and develop remediation and prevention strategies. 

\subsection{Motivating Example}
\par
As shown in Figure~\ref{fig:dependency graph}, this is a typical data leakage attack. The attacker exploited a vulnerability in \texttt{apache2}  and downloaded the malicious artifact \texttt{gather.sh}. After executing the malware, the attacker collected sensitive data from the target host and saved it in the form of the file \texttt{leaked.vm2}. After using \texttt{gpg} to compress \texttt{leaked.vm2} into the file \texttt{leaked}, the attacker transferred \texttt{leaked} to the C2 server \texttt{192.168.2.3:xx} via process \texttt{ssh}.
\par
In Figure~\ref{fig:dependency graph}, the black dashed box denotes the \textit{backtracking graph} obtained by performing backward causality analysis~\cite{king2003backtracking}, which includes \textbf{all events causal-related to the alert}. The blue dashed box denotes the \textit{suspicious semantic graph} obtained by using suspicious semantic transfer, which includes \textbf{all events  suspicious semantic-related to the alert}. The red dashed box denotes the \textit{critical component graph} obtained by analyzing the path-level contextual semantics, which includes \textbf{all events attack-related to the alert}.

\par
Obviously, the number of attack-related critical events ($\sim$ 22) is a drop in the ocean compared to the number of causal-related non-critical events ($ \sim $ 200,000). This turns the attack investigation into a needle-in-a-haystack process, making it challenging for analysts to complete the investigation in the optimal time (600s)~\cite{lateral-movement}. However, existing technique such as  DEPIMPACT~\cite{fang2022depimpact}, as shown in Section~\ref{sec:5_evaluation}, has exhibited poor performance. It requires 6,464s ($\gg$ 600s) to generate dependency graphs with 2,473 false positives on average. In addition, it needs to load raw audit logs, resulting in an endless memory overhead. Therefore, we need an attack investigation system with low false positives, low latency and low overhead.

\section{Overview}
\label{sec:3_overview}

\par
\subsection{Threat Model}
\label{sec:3.2_threatmodel}
\par
First, we assume that the event logs and digital signatures are credible, similar to previous work~\cite{gao2021enabling, gao2018saql, gao2018aiql, hassan2019nodoze, king2003backtracking, liu2018priortracker, hassan2020rapsheet, fang2022depimpact}. In addition, events related to the attack did not occur before the logs were processed.
\par
Second, we assume that the attacker is external to the system and carries out their attack remotely. This may involve exploiting vulnerabilities within the system or employing social engineering tactics to convince a user to download and run a file containing malicious code. Therefore, we do not support side-channel attacks and insider attacks where the attacker has a legitimate way to access the machine without going through them.
\par
Third, we exclude mimicry attacks~\cite{wagner2002mimicry} from consideration in our threat model. These attacks are designed to evade intrusion detection systems by creating a seemingly benign chain of events within an enterprise environment. Existing intrusion detection systems~\cite{bishop2005introduction, kruegel2004intrusion, insider-threat-monitoring-software} often rely on heuristics or analysis of individual events, making them vulnerable to such attacks. While detecting mimicry attacks is a limitation of current detection systems, it falls outside the scope of our work. Our focus is on identifying relevant events of alert generated by the detection system as contextual information to investigate the attack.

\subsection{Our Approach}
In this section, we describe the architecture of \sn{} shown in Figure~\ref{fig:sysarchi}. Given a POI event, \sn{} can automatically identify the critical component of the dependency graph. \sn{} consists of two phases: (\uppercase\expandafter{\romannumeral1}) suspicious semantic graph construction (SSGC) and (\uppercase\expandafter{\romannumeral2}) path-level contextual analysis (PCA).
\par
    In Phase \uppercase\expandafter{\romannumeral1}, \sn{} makes use of mature auditing systems~\cite{etw, auditd, lttng, sysdig, redhat} to access kernel-level streaming logs and process them into specific data structures. Then \sn{} proposes a \textit{suspicious semantic transfer rule and storage strategy} to maintain the suspicious entity list and related event table with low memory overhead. Given a POI event, \sn{} can construct the \textit{suspicious semantic graph (SSG)} in real-time. 
\par
In Phase \uppercase\expandafter{\romannumeral2}, \sn{} first performs edge compaction on the suspicious semantic graph. Then \sn{} proposes a \textit{suspicious flow path extraction algorithm} to identify possible propagation paths of the data/control flow in the suspicious semantic graph (i.e., \textit{suspicious flow paths}). Next, \sn{} performs \textit{path-level contextual analysis}, scores each suspicious flow path, and determines how relevant the path is to the POI event. Finally, \sn{} filters out all events that only exist in irrelevant paths from suspicious semantic graph to generate the \textit{critical component graph (CCG)} as the output.
\section{System Design}
\label{sec:4_sysdesign}

\begin{figure*}[h!t]
\centering
\includegraphics[width=\linewidth]{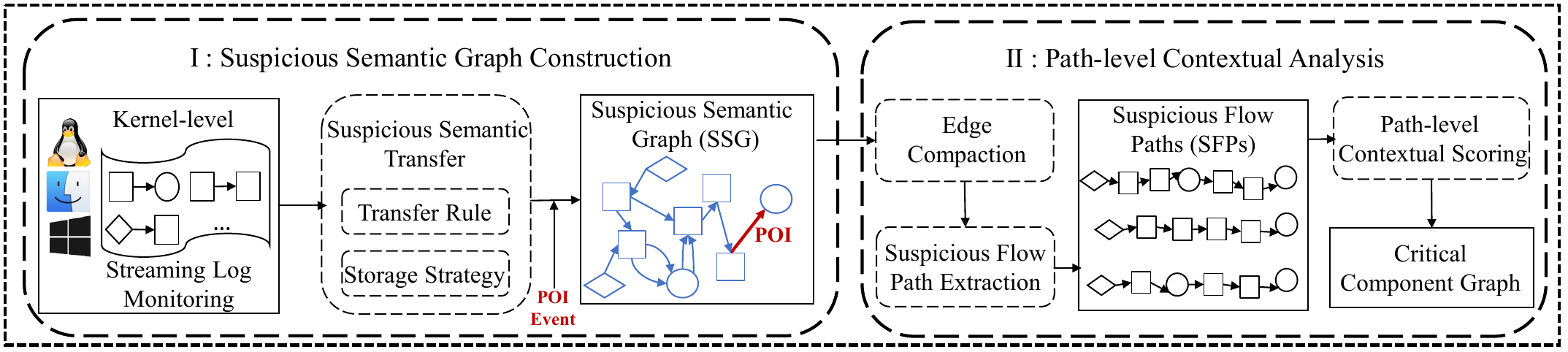}
\setlength{\abovecaptionskip}{-0.2cm}
\caption{Architecture of \sn{}.}
\label{fig:sysarchi}
\vspace{-0.2cm}
\end{figure*}

\par
In this section, we describe the design details of each phase of \sn{}. As shown in Figure~\ref{fig:sysarchi}, \sn{} is a two-phase framework (i.e., constructing suspicious semantic graph and performing path-level contextual analysis) for mitigating the dependency explosion problem. 

\subsection{Goal and Key Insight}
\label{sec:4.1_goals}
\subsubsection{Suspicious Semantic Graph Construction}
\label{sec:4.1.1_goals_ssg}
\ \\
\textbf{Goal.} Given an alert point, current investigation techniques~\cite{hassan2019nodoze, hassan2020swift, liu2018priortracker, fang2022depimpact, ma2016protracer} store audit logs in memory (\textbf{high overhead}) and construct a backtracking graph~\cite{king2003backtracking} from the alert point.  However, it usually includes numerous events that are impossible to result in an attack (\textbf{high false positives}), such as reads to read-only files, and interactions with benign processes. Additionally, it takes time to identify related events by going through these logs (\textbf{high latency}). In summary, the backtracking graph gives rise to the problems of high memory overhead, high time overhead, and high false positives in existing approaches. Therefore, \textit{we aim to construct a suspicious semantic graph with low memory overhead in real-time, which is smaller in size than the backtracking graph but contains all attack-related events.}
\par
\textbf{Key Insight.} To achieve the aforementioned goal, there are two key insights upon which we rely. 
(1) Suspicious semantics are introduced externally, i.e., attack is implemented remotely, as defined in Section~\ref{sec:3.2_threatmodel}.
(2) Suspicious semantics propagate between entities, i.e., suspicious entities transmit the suspicious semantics to non-suspicious entities via interaction. 
\par
Based on the above insights, we present a state-based framework to achieve the goal, which includes Section~\ref{sec:4.2_streaming_log} Streaming Log Monitoring and Section~\ref{sec:4.3_susSemanTrans} Suspicious Semantic Transfer.

\subsubsection{Path-level Contextual Analysis}
\label{sec:4.1.2_goals_sfp}
\ \\
\textbf{Goal.} Once the POI event is identified, we construct the corresponding suspicious semantic graph. This suspicious semantic graph consists of all events semantically related to the POI event and contains all attack-related events (i.e., critical events). As shown in Section~\ref{sec:5_evaluation}, the SSG ($\sim$ 417 edges) is 545 $times$ smaller than the backtracking graph ($\sim$ 227,589 edges) but 3.7 $times$ larger than the critical component graph ($\sim$ 113 edges). This suggests that there are still many false positives in the suspicious semantics graph. Therefore, \textit{we aim to filter out the events that are contextually irrelevant to the POI event in the suspicious semantics graph by performing path-level contextual analysis.} By mitigating the dependency explosion problem for the second time, we obtain a \textit{critical component graph} to assist analysts in conducting attack investigation.
\par
\textbf{Key Insight.} To achieve path-level contextual analysis, we rely primarily on the following two key insights : (1) Only by evaluating data/control flow paths as a whole we can determine whether they have an impact on the POI event. In other words, we cannot determine whether an event has impacted a POI event in isolation (i.e., at the event-level)~\cite{hassan2019nodoze, fang2022depimpact}, but rather in context (i.e., at the path-level). (2) Quantifying the degree of impact requires consideration of the properties and neighboring relationships of events. 
\par
Based on the above insights, we propose a path-level contextual analysis mechanism consisting of Section~\ref{sec:4.4_edgeCompat} Edge Compaction, Section~\ref{sec:4.5_bfs} Suspicious Flow
Path Extraction and Section~\ref{sec:4.6_pathScroing} Path-level Contextual Scoring.

\subsection{Streaming Log Monitoring}
\label{sec:4.2_streaming_log}
\par
\sn{} makes use of mature auditing systems~\cite{etw, auditd, lttng, sysdig, redhat} to access kernel-level logs and obtain the required data. At the entity level, \sn{} focuses on three entity types: file, process, and socket. 
To differentiate, \sn{} needs to construct unique identifiers for all entities. For the file, \sn{} records the absolute path as the unique identifier. For the process, \sn{} concatenates the PID and name as the unique identifier. For the socket, \sn{} constructs the 4-tuple (\textit{\textless srcip, srcport, dstip, dstport\textgreater}) as the unique identifier. At the event level, \sn{} focuses on three event types: process interactions, file IO events, and network IO events.
To the best of our knowledge, existing auditing systems are rich in semantics and meet the data requirements of \sn{}.

\begin{table*}[]
\centering
\caption{Suspicious Semantic Transfer Rule.}
\label{tab:sus_semantic_rules}
\resizebox{\textwidth}{!}{%
\begin{tabular}{c|c|c|c}
\hline
\textbf{Event Type} & \textbf{Subject} & \textbf{Object} & \textbf{Description}                                                       \\ \hline
Recvfrom            & Socket           & Process         & A process receives data from the network, the process becomes suspicious. \\
Sendto              & Process          & Socket          & A suspicious process sends data to the network.                            \\
Read                & File             & Process         & A process reads a suspicious file, the process becomes suspicious.      \\
Write               & Process          & File            & A suspicious process writes a file, the file is suspicious.              \\
Execve/Clone        & Process          & Process         & A process is started by a suspicious process, the process is suspicious.   \\ \hline
\end{tabular}%
}
\end{table*}

\subsection{Suspicious Semantic Transfer}
\label{sec:4.3_susSemanTrans}
The letter P in APT stands for persistence, which means that an attacker can lurk for a long time until achieves the goal. To support real-time investigation and long-term monitoring, \sn{} utilizes a state-based structure and suspicious semantic transfer rule to record state changes and associated events for each entity. We next describe the specific data structure and transfer rule in turn.
\subsubsection{Data Structure}
\ \\
For any entity $ v \in V$, \sn{} represents it as a triple $<U,\ T_y,\ S>$. $U$ is the unique identifier of the entity, the construction of $U$ is described in Section~\ref{sec:4.2_streaming_log}. $T_y$ denotes the type of the entity and S denotes the state of the entity. When $S$ is 0 it means that the entity is not suspicious, and $S$ is 1 it means that the entity is suspicious. Note that the file and process have their $S$ initialized to 0 when they are created, and the socket has their $S$ initialized to 1 when it is created, i.e., we default to all sockets that are not in the whitelist being suspicious (\textbf{Key Insight (1)} in Section~\ref{sec:4.1.1_goals_ssg}).
\par
For any event $ e \in E$, \sn{} represents it as a quintuple $<U_{s}, U_{o}, O, T_{i}, D>$. $U_s$ and $U_o$ are unique identifiers for the subject and object of $e$, respectively. $O$ denotes the type of $e$, $T_i$ denotes the time when $e$ occurred, and $D$ denotes the data flow amount of $e$. Note that \sn{} is based on the direction of the data flow and control flow to determine the location of the subject and object. For example, when $O = Read$, the data flow is from the file to the process, so the file is the subject and the process is the object. When $O = Write$, the data flow is from the process to the file, so the process is the subject and the file is the object.

\subsubsection{Transfer Rule}
\label{sec:4.3.2_susRules}
\begin{figure*}[h!t]
\centering
\includegraphics[width=\linewidth]{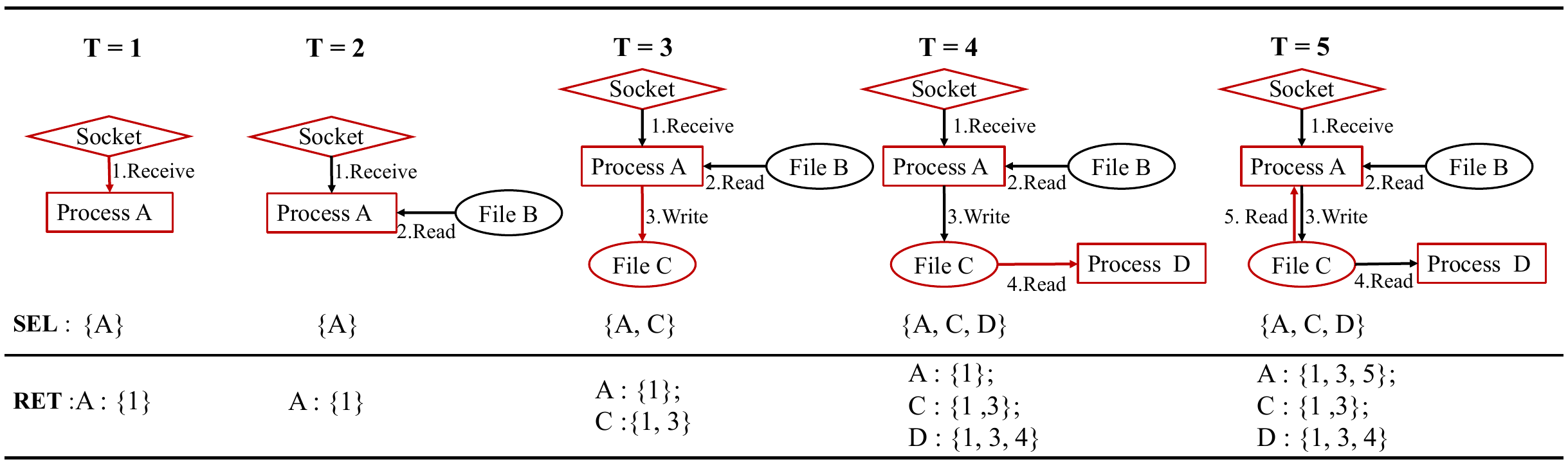}
\setlength{\abovecaptionskip}{-0.3cm}
\caption{\centering An Example of Suspicious Semantic Transfer. The red solid line indicates that the entity carries suspicious semantic. \textbf{SEL} is short for Suspicious Entity List and \textbf{RET} is short for Relevant Event Table.}
\label{fig:semanTransExample}
\end{figure*}

\ \\
Based on the idea of semantic transfer (\textbf{Key Insight} (2) in Section~\ref{sec:4.1.1_goals_ssg}), \sn{} constructs a set of predefined rules to process streaming logs and identify entity states in real-time. As shown in Table~\ref{tab:sus_semantic_rules}, each rule is a quadruple: $<O,\ T_{s},\ T_{o},\ D>$. $O$ is the type of event, $T_{s}$ and $T_{o}$ are the entity types of the subject and object respectively, and $D$ is a description of the rule. From Table~\ref{tab:sus_semantic_rules} we can see that the subject is able to transfer suspicious semantics to the object via a specific event, which is referred to as the "suspicious semantics transfer rule". As shown in Figure~\ref{fig:semanTransExample}, $T$ denotes the moment, red entities denote suspicious entities, and red straight arrows denote suspicious semantic transfer. When $T = 3,$ a suspicious process ($process\ A$) writes data to a file ($file\ C$), which in turn carries the suspicious semantic. When $T = 4$, the suspicious file is read by another process ($process\ D$), which then carries the suspicious semantic. Conversely, if an entity has no suspicious semantic, any event involving this entity as a subject will not propagate suspicious semantic. For example, when $T = 2$, a file ($file\ B$) is read by the suspicious entity ($process\ A$), but there is no propagation of the suspicious semantic.
\par
As shown in lines 5 to 11 in Algorithm~\ref{alg:ssg_cons}, \sn{} processes the streaming logs, analyses data flows, and determines whether the entity state transitions. First, \sn{} accesses the event $e = <U_s, U_o, O, T_i, D>$ and constructs the subject and object $u$, $v$ corresponding to that event. Then, \sn{} determines whether the subject $u$ is a socket or exists in \textit{suspicious entity list} (SEL, see Section~\ref{sec:4.3.3_storage} for detailed definition). Finally, as soon as one of these two conditions is met, \sn{} will mark the object $v$ as suspicious and add it to SEL.

\begin{algorithm}[!h]
    \caption{Suspicious Semantic Graph Construction}
    \label{alg:ssg_cons}
 \begin{algorithmic}[1]
    \REQUIRE  
        \STATE (1) Streaming logs in chronological order; 
        \STATE (2) Suspicious Entity List (SEL); 
        \STATE (3) Related Event Table (RET);
        \STATE (4) POI event $p$;
     \ENSURE 
        Suspicious semantic graph for POI event $p$;

        \FOR{$e \in Streaming\ \  logs$}
        \STATE Construct $u, v$ from $e$ where $u_{U} = e_{U_{s}}, v_{U} = e_{U_{o}}$ 
            \IF {$u_{S} == 0 \  \textbf{and} \ \nexists \ \   u_{U} \in SEL $}
                \STATE continue;
            \ELSE
            \STATE $v_{S} = 1$;
            \IF { $\nexists \ \   v_{U} \in SEL$  }
                \STATE SEL.append($v_{U}$);
            \ENDIF
            \STATE Add \{$v_{U} :$ RET[$u_{U}$] + e\} to RET;
            \ENDIF
            \IF{ $\exists \ p_{U_{o}} \in  SEL$}
        \STATE \RETURN graphConstruct(RET[$p_{U_{o}}$])
        \ENDIF
    \ENDFOR

    \end{algorithmic}
\end{algorithm}

\begin{algorithm}[h]  
  \caption{Suspicious Flow Path Extraction}  
  \label{alg:sfp_extra}  
  \begin{algorithmic}[1]
    \REQUIRE  
        \STATE (1) Suspicious Semantic Graph $G$; 
        \STATE (2) POI Event $p$; 
        \STATE (3) $Q$ and $V$ for the queue structure, $T$ for the tree structure;

     \ENSURE
        Suspicious flow paths;
    \STATE $Q.add(p)$
    \STATE $T.creatNode(p)$
    \WHILE{$Q.num \ne 0$}
        \STATE $e = Q.pop()$
        \STATE $V.add(e)$
        \FOR{$ie\  \in\  G.inEdges(e_{U_{s}})$}
            \IF{$ie_{T_{i}} < e_{T_{i}}\  \ \textbf{and} \  \ ie \notin Q \ \textbf{and} \  ie \notin V$}
              \STATE $Q.add(ie)$
              \STATE $T.creatNode(ie)$
              \STATE $T.creatEdge(e,\  ie)$
            \ENDIF
        \ENDFOR
    \ENDWHILE
    \RETURN $T.allPaths()$
  \end{algorithmic}  
\end{algorithm}

\subsubsection{Storage Strategy}
\label{sec:4.3.3_storage}
\ \\
\sn{} designs two data structures to enable efficient storage of relevant data and real-time construction of the suspicious semantic graph. Specifically, \sn{} designs a \textit{Suspicious Entity List (SEL)} and a \textit{Related Event Table (RET)}, as defined below.
\par
\textbf{Suspicious Entity List:} A list that maintains all entities with suspicious semantics (possibly related to attacks). As shown in Figure~\ref{fig:semanTransExample}, when $T=1$, the data flow passes from the suspicious socket to process $A$ (suspicious semantic transfer), so \sn{} adds entity $A$ to SEL. When $T=2$, there is no suspicious semantic transfer, so SEL is not changed. When $T=5$, the data flow passes from suspicious file $C$ to process $A$, but entity $A$ is already in SEL, so SEL is not changed.
\par
\textbf{Related Event Table:} A table that holds all the related events corresponding to all suspicious entities. \textit{The related events of a suspicious entity refer to the set of all data flows and control flows that lead to this entity's semantic change.} Specifically, \sn{} will maintain a separate set of related events in RET for all suspicious entities. Whenever an event $e=<U_{s},\ U_{o},\ O,\ T_{i},\ D>$ that satisfies the suspicious semantic transfer rule occurs, \sn{} will stitch the related events of $U_{s}$ with event $e$ and use it as the related events of $U_{o}$ to update RET. As shown in Section~\ref{sec:5.3_efficient}, the size of the RET is much smaller than the raw audit logs and the time of read in is negligible. \sn{} can construct a suspicious semantic graph related to the POI event in real-time.
\par
As shown in Figure~\ref{fig:semanTransExample}, when $T=1$, \sn{} adds to RET with $A:\ \{1\}$, indicating that the related event of the suspicious entity $A$ is $\{1\}$. When $T=4$, \sn{} adds to RET with $D:\ \{1, 3, 4\}$, stitched from the related events $\{1, 3\}$ of subject $C$ and the current event $\{4\}$. When $T=5$, \sn{} updates RET with $A:\ \{1, 3, 5\}$, stitched from the related events $\{1, 3\}$ of subject $C$ and the current event $\{5\}$.
\par
In order to speed up the consumption of log streams, \sn{} keeps the whole SEL in memory to determine the entity states and save suspicious entities in real-time. 
In contrast, inspired by the CPU architecture, \sn{} keeps only some of the high-modification (frequent growth in a short period) RETs in memory and stores other low-modification RETs in the hard disk. According to our experimental results (see Section~\ref{sec:5.3_efficient} for detail), the memory overhead of \sn{} is 30MB on average, and there is no problem with high memory overhead. \textit{Note that we default sockets to suspicious entities, so only entities of file type and process type are saved in SEL.}
\par
In summary, \sn{} will use these two data structures to enable efficient storage of the necessary data and real-time construction of the suspicious semantic graph. As shown in lines 11 to 17 of Algorithm~\ref{alg:ssg_cons}, \sn{} will add the object $v_U$ to the SEL for any event $e=<U_{s},\ U_{o},\ O,\ T_{i},\ D>$ that satisfies the semantic transfer rule. In addition, \sn{} stitches the related events of $U_{s}$ with event $e$ and uses it as the related events of $U_{o}$ to update RET. Finally, for any given POI event, \sn{} is able to extract all relevant events for the object of that POI event from the RET in real-time. \sn{} then uses a simple graph construction algorithm, which extracts entities from entities as nodes and events as edges, to construct a suspicious semantic graph associated with the POI event.


\subsection{Edge Compaction}
\label{sec:4.4_edgeCompat}
A suspicious semantic graph often contains multiple parallel edges between two nodes. This is because operating systems typically complete read/write tasks (e.g., file read/write) by proportionally allocating data to multiple system calls. Inspired by recent work for graph reduction~\cite{xu2016cpr}, \sn{} merges the edges between two nodes if the time difference between them is less than a given threshold. We ultimately chose 10 seconds as it demonstrates reasonable results in terms of various system calls, such as file transfers and network connections.

\begin{figure*}[h!t]
\centering
\includegraphics[width=\linewidth]{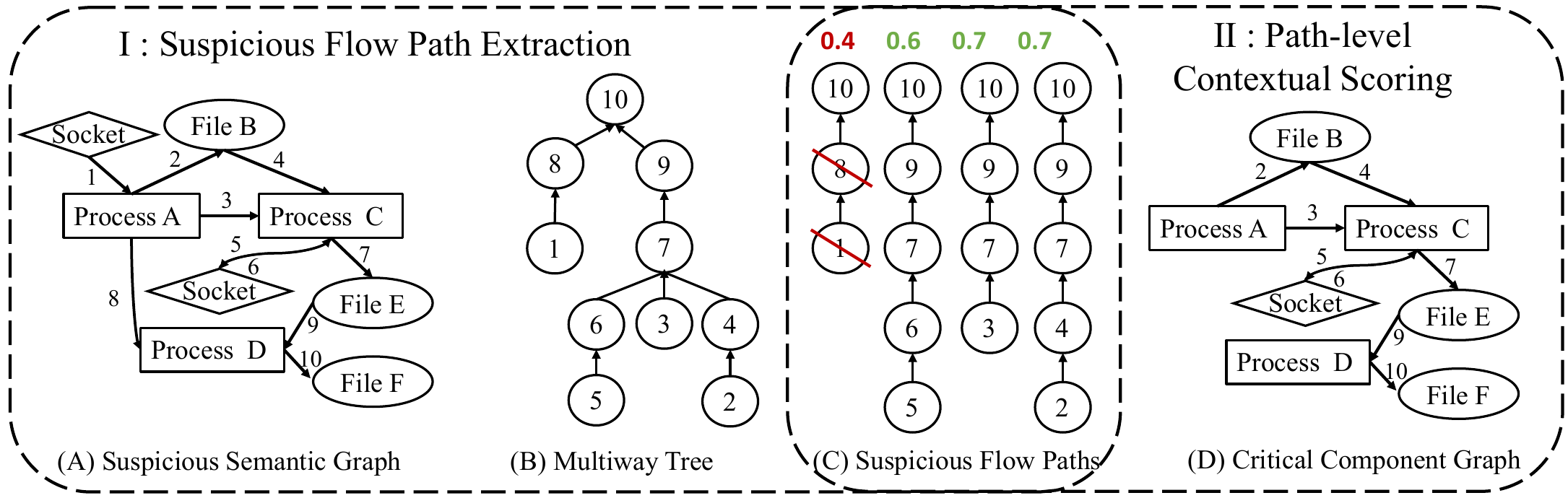}
\setlength{\abovecaptionskip}{-0.2cm}
\caption{Suspicious flow path extraction and path-level contextual scoring.}
\label{fig:sfp_extra}
\vspace{-0.2cm}
\end{figure*}

\subsection{Suspicious Flow Path Extraction}
\label{sec:4.5_bfs}
\par
In order to perform path-level contextual analysis, it is first necessary to identify possible propagation paths of the data/control flow in the suspicious semantic graph (i.e., \textit{suspicious flow paths}). \sn{} proposes a \textit{suspicious flow path extraction algorithm} that can efficiently handle complex graph structures. In brief, as shown in Figure~\ref{fig:sfp_extra}, \sn{} transforms the suspicious semantic graph into a multiway tree and then traverses it to obtain all suspicious flow paths.

\par
Specifically, as shown in lines 1 to 5 of Algorithm~\ref{alg:sfp_extra}, $Q$ and $V$ are the queue structures, where $Q$ holds the events to be traversed and $V$ holds the events that have been traversed. $T$ is the multiway tree structure, which holds the topological information. As shown in lines 6 to 9 of Algorithm~\ref{alg:sfp_extra}, \sn{} traverses event $e$, identifying all incoming edge $ies$ ($ies = G.inEdges(e_{U_{s}})$). As shown in lines 10-13 of Algorithm~\ref{alg:sfp_extra}, \sn{} determines that the incoming edge $ie$ ($ie \in ies$) occurred earlier than event $e$ and has not been traversed ($ie \notin V$), then creates node $ie$ in the multiway tree $T$ and the parent of that node is $e$. Finally, \sn{} traverses the multiway tree $T$ to obtain all paths from the root node to the leaf nodes, which are output as suspicious flow paths.
\par
The suspicious flow path extraction algorithm takes into account the timeliness and directionality of the data/control flow and is able to handle the complex graph structure efficiently, as demonstrated in Section~\ref{sec:5.3_efficient}, where \sn{} extracts over 140 suspicious flow paths in one second on average. Finally, it is important to note that events exist as nodes in the multiway tree and suspicious flow paths, as shown in Figure~\ref{fig:sfp_extra}.

\subsection{Path-level Contextual Scoring}
\label{sec:4.6_pathScroing}
After extracting the suspicious flow paths, \sn{} needs to perform contextual analysis at the path-level to quantify the degree of influence of the entire path on the POI event (\textbf{Key Insight (1)} in Section~\ref{sec:4.1.2_goals_sfp}). Furthermore, the degree of impact between events is determined by the event attributes and the neighboring relationships between events (\textbf{Key Insight (2)} in Section~\ref{sec:4.1.2_goals_sfp}).
\par
For each suspicious flow path $p$, \sn{} calculates the $PathScore$ using the following equation:
{\setlength\abovedisplayskip{0.1cm}
\setlength\belowdisplayskip{0.1cm}
\begin{equation}
PathScore = \sum_{E}^{e} EventScore(e)\  / \ Len(p)
\end{equation}}
where $e$ denotes an event and $E$ denotes the set of all events contained in the path $(e \in E)$. $EventScore$ denotes the degree of impact of event $e$ on the parent node, as defined later. $Len(p)$ denotes the number of events in the path and is used to normalize the $PathScore$.
\par
\sn{} calculates the $EventScore$ using the following equation:
{\setlength\abovedisplayskip{0.1cm}
\setlength\belowdisplayskip{0.1cm}
\begin{equation}
EventScore = \alpha \frac{Impact(e, f)}{\sum_{child(f)}^{s} Impact(s, f)}\ ,\ f=parent(e)
\end{equation}}

{\setlength\abovedisplayskip{0.1cm}
\setlength\belowdisplayskip{0.1cm}
\begin{equation}
\alpha  = 1 + \frac{len(child(f) - 1)}{C} \label{euqation3}
\end{equation}}
where $parent(e)$ denotes the parent of event $e$, and $child(f)$ denotes all the children of event $f$ in the multiway tree. $Impact(e, f)$ denotes the degree of impact that event $e$ exerts on event $f$, as defined later. $\alpha$ is an inflation factor to mitigate the problem of decreasing relative impact due to triage (a parent node with multiple children). As shown in Equation~\ref{euqation3}, $\alpha$ is controlled by the super parameter $C$ and the number of child nodes. It is negatively correlated with $C$ and positively correlated with the number of child nodes.
\par
\sn{} picks two features (i.e., data flow amount and time), to calculate the $Impact$ using the following equation:
{\setlength\abovedisplayskip{0.1cm}
\setlength\belowdisplayskip{0.1cm}
\begin{equation}
\begin{aligned}
Impact(e1, e2) = CS(Nor(e1_{D}, e1_{T_{i}}),  Nor(e2_{D}, e2_{T_{i}}))
\end{aligned}
\end{equation}
}
where $e1$ and $e2$ are two events and $e2$ is the parent node of $e1$ (i.e., $e2 = parent(e1)$). $e_D$ and $e_T$ denote the data flow and occurrence time of event $e$, respectively. $Nor(\cdot)$ denotes normalization, which removes differences in $e_D$ and $e_T$ on the scale. $CS(\cdot)$ denotes the computation of cosine similarity. 
\par
Intuitively, we assume that if the data flow amount between parent node and child node is similar, then there is a causal relation between them (e.g., a process reads 526 bytes from the network and then immediately writes 526 bytes to a file, which may be the same content). Similarly, if the timestamps are similar, then there is a causal relation between the events since we think that the exploitation is automated and its steps quickly follow each other.

\par
\sn{} will iteratively calculate the scores of all suspicious flow paths and consider the path whose score is below a threshold $T$ as an $irrelevant \ path$. Then, \sn{} filters out events that only exist in the $irrelevant \ path$ and outputs the retained part as a \textit{critical component graph} to help analysts in attack investigation.
\section{Evaluation}
\label{sec:5_evaluation} 
In this section, we first present the evaluation preparation, including the characteristics of the dataset, the obtaining of ground truth, and the setting of evaluation metrics. We then evaluate the effectiveness and efficiency of each component separately. In summary, we aim to answer the following questions:
\begin{itemize}
\item{\textbf{RQ1:} How effective is \sn{} in attack investigation?}
\item{\textbf{RQ2:} How efficient is \sn{} in attack investigation?} 
\item{\textbf{RQ3:} How sensitive is \sn{} in parameter selection?}
\end{itemize}

\subsection{Evaluation Preparation}
\par
We deploy our implementation of \sn{} on a computer with Intel (R) Core (TM) i9-10900K CPU @ 3.70GHz and 64GB memory. \sn{} processes streaming logs from the auditing systems Sysdig~\cite{sysdig} and SPADE~\cite{gehani2012spade}, extracts information in the format as described in Section~\ref{sec:2.1_dataformat}, and runs continuously in a low-overhead state. 



\subsubsection{Attack Dataset}
\ \\
We evaluate the effectiveness of \sn{} in revealing attack sequences on a dataset with over 150 million system audit logs. As shown in Table~\ref{tab:statis_Dataset}, this dataset contains 15 attack cases (10 simulated attacks and 5 DARPA attacks), and is provided by DEPIMPACT~\cite{fang2022depimpact}. The simulated attacks consist of 7 (rows 2 to 8) single-host attacks based on common exploits~\cite{explot-database, ma2016protracer, xu2016cpr, kwon2018mci} and 3 (rows 9 to 11) multi-host attacks based on Cyber Kill Chain~\cite{cyber-kill-chain} and CVE reports~\cite{explot-database}. The simulated attacks utilized deployed hosts with 12 active users and hundreds of processes, daily tasks such as file manipulation, text editing, and software development were carried out to simulate real-world usage. We detail these 10 simulated attacks in Appendix~\ref{sec:appendix_common-7} and Appendix~\ref{sec:appendix_multi-3}. The DARPA dataset contains 5 host attacks (rows 12 to 16), which was done by two teams (FiveDirections and Theia), and differed in terms of target systems (Windows, Linux) and vulnerability exploits (pine backdoor, firefox backdoor, and browser extension).
\par
Table~\ref{tab:statis_Dataset} shows the statistics of the generated dependency graphs for all attacks. Column ``Attack'' indicates the name of the attack case. Columns ``\# V'' and ``\# E'' indicate the number of nodes and edges of the backtracking graphs after performing causality analysis~\cite{king2003backtracking} from POI events. Column ``\# CE'' shows the number of critical events (related to the attack), which we explain in detail below.

\begin{table}[]
\small
\centering
\caption{\centering The statistics of dependency graphs generated for all the 15 attacks.}
\label{tab:statis_Dataset}
\resizebox{0.45\textwidth}{!}{%
\begin{tabular}{crrr}
\hline
\textbf{Attack}      & \textbf{\# V} & \textbf{\# E} & \textbf{\# CE} \\ \hline
Wget Executable      & 78            & 349           & 16             \\
Illegal Storage      & 2,277         & 34,367        & 7              \\
Illegal Storage2     & 9,345         & 290,933       & 7              \\
Hide File            & 23,110        & 459,514       & 10             \\
Steal Information    & 23,153        & 495,570       & 7              \\
Backdoor Download   & 1,411         & 12,354        & 12             \\
Annoying Server User & 114           & 585           & 15             \\
Shellshock           & 1,706         & 42,918        & 36             \\
Dataleak             & 1,863         & 20,807        & 25             \\
VPN Filter           & 2,436         & 39,332        & 29             \\
Five Dir Case 1      & 259           & 473           & 8              \\
Five Dir Case 3      & 6,109         & 83,154        & 9              \\
Theia Case 1         & 175,196       & 794,341       & 8              \\
Theia Case 3         & 281,001       & 1,137,829     & 8              \\
Theia Case 5         & 245           & 1,309         & 5              \\ \hline
\textbf{Avg}         & 35,220.20     & 227,589.00     & 13.47          \\ \hline
\end{tabular}%
}
\end{table}

\begin{table*}[]
\scriptsize
\centering
\caption{Performance of dependency graphs generated by different technique. SSGC and PCA are the components of \sn{}.}
\label{tab:diff_tech_per}
\resizebox{\textwidth}{!}{%
\begin{tabular}{c|ccc|ccc|ccc|ccc|cccc}
\hline
\multirow{2}{*}{\textbf{Attack}} & \multicolumn{3}{c|}{\textbf{SLEUTH}} & \multicolumn{3}{c|}{\textbf{NODOZE}} & \multicolumn{3}{c|}{\textbf{DEPIMPACT}} & \multicolumn{3}{c|}{\textbf{SSGC}} & \multicolumn{4}{c}{\textbf{SSGC+PCA}}                  \\
                                 & FP         & FN         & \# E       & FP         & FN         & \# E       & FP          & FN          & \# E        & FP         & FN        & \# E      & FP             & FN            & \# E         & \# SFP \\ \hline
Wget Executable                  & 68         & 7          & 77         & 78         & 0          & 94         & 32          & 0           & 48          & 3          & 0         & 19        & 1              & 0             & 17           & 5      \\
Illegal Storage                  & 2189       & 5          & 2191       & 5686       & 1          & 5694       & 2625        & 0           & 2632        & 172        & 0         & 179       & 47             & 0             & 54           & 85     \\
Illegal Storage2                 & 2072       & 3          & 2076       & 11959      & 1          & 11967      & 1255        & 0           & 1262        & 986        & 0         & 993       & 252            & 0             & 259          & 522    \\
Hide File                        & 4558       & 5          & 178703     & 38919      & 2          & 38931      & 14982       & 0           & 14992       & 1590       & 0         & 1600      & 356            & 0             & 366          & 833    \\
Steal Information                & 3972       & 3          & 179202     & 21114      & 1          & 21122      & 15774       & 0           & 15781       & 1654       & 0         & 1661      & 453            & 0             & 460          & 868    \\
Backdoor Download                & 1392       & 6          & 9198       & 232        & 1          & 245        & 2625        & 0           & 2637        & 42         & 0         & 54        & 36             & 0             & 48           & 20     \\
Annoying Server User             & 2          & 13         & 281        & 88         & 2          & 105        & 39          & 0           & 54          & 3          & 0         & 18        & 1              & 0             & 16           & 6      \\
Shellshock                       & 672        & 9          & 4299       & 1007       & 4          & 1047       & 161         & 4           & 197         & 87         & 0         & 123       & 9              & 0             & 45           & 40     \\
Dataleak                         & 622        & 7          & 4279       & 597        & 7          & 629        & 44          & 3           & 69          & 16         & 0         & 47        & 68             & 0             & 83           & 17     \\
VPN Filter                       & 722        & 8          & 5342       & 310        & 5          & 344        & 189         & 2           & 218         & 86         & 0         & 115       & 72             & 0             & 101          & 37     \\
Five Dir Case 1                  & 97         & 4          & 237        & 291        & 1          & 300        & 17          & 0           & 25          & 8          & 0         & 16        & 3              & 0             & 11           & 11     \\
Five Dir Case 3                  & 277        & 5          & 37058      & 717        & 1          & 727        & 171         & 0           & 180         & 82         & 0         & 91        & 31             & 0             & 40           & 22     \\
Theia Case 1                     & 453        & 6          & 264107     & 92091      & 2          & 92101      & 689         & 1           & 697         & 494        & 0         & 502       & 98             & 0             & 106          & 374    \\
Theia Case 3                     & 231        & 4          & 493626     & 7369       & 1          & 7378       & 1074        & 1           & 1082        & 820        & 0         & 828       & 61             & 0             & 129          & 598    \\
Theia Case 5                     & 2          & 1          & 845        & 395        & 1          & 401        & 3           & 0           & 8           & 9          & 0         & 14        & 2              & 0             & 7            & 15     \\ \hline
\textbf{Avg FP/FN/\# E}          & 1154       & 5.73       & 1163       & 16682      & 2.00       & 16696      & 2473        & 0.73        & 2487        & 403        & 0.00      & 417       & \textbf{99}    & \textbf{0.00} & \textbf{113} & 230    \\ \hline
\textbf{Avg FPR/FNR}($10^{-2}$)             & 0.507      & 42.54      & /          & 7.329      & 14.85      & /          & 1.087       & 5.419       & /           & 0.177      & 0         & /         & \textbf{0.043} & \textbf{0}    & /            & /      \\ \hline
\end{tabular}%
}
\begin{tablenotes}
    \centering
    \footnotesize
    \item $\dagger $ \textbf{SSGC}:Suspicious Semantic Graph Construction. \textbf{PCA}:Path-level Contextual Analysis. \textbf{SFP}:Suspicious Flow Path. 
\end{tablenotes}
\end{table*}

\subsubsection{Obtaining Ground Truth}
\label{sec:5.1.2_groundtruth}
\ \\
In order to evaluate the performance of \sn{}, we need to specify the ground truth for all attack cases (i.e., identify all critical events). Specifically, we analyzed the targets of each attack case and determined the corresponding POI events from massive logs. We then conducted  back-propagation~\cite{king2003backtracking} from POI events to obtain backtracking graphs and searched for critical events within them. Finally, we manually ascertained the critical events based on Indicators of Compromise (e.g., file names and malware names) and attack steps (e.g., download then execution), as shown in Appendix~\ref{sec:appendix_attacksample}.

\par
\textbf{Evaluation Metrics.}
First, we measure \textit{false positives (\textbf{FP})} and \textit{false negatives (\textbf{FN})}. False positives refer to those edges that \sn{} identifies as critical but are not, while false negatives refer to those edges that \sn{} identifies as irrelevant but are critical.  Then we compute the false positive rate $FPR=FP/E_{total}$ and false negative rate $FNR=FN/E_{c}$, where $E_{total}$ represents the number of edges and $E_{c}$ represents the number of critical edges, respectively.

\subsection{RQ1 : How effective is \sn{} in attack investigation ?}
\label{sec:5.2_effective}
There has been a lot of graph-based related work on attack investigation~\cite{milajerdi2019holmes, pei2016hercule, hossain2017sleuth, zeng2021watson,xu2016cpr, fang2022depimpact, hassan2019nodoze, hassan2020rapsheet}. However, HOLMES\cite{milajerdi2019holmes} and RapSheet\cite{hassan2020rapsheet} rely solely on defined TTP-like (Tactics, Techniques, and Procedures) rules for detection and investigation. Such approaches suffer from heavy reliance on manual efforts and cannot effectively address zero-day vulnerabilities. HERCULE\cite{pei2016hercule} and WATSON\cite{zeng2021watson} address attack investigation problems by discovering communities (i.e., behavioral abstractions) on the provenance graph. Their purpose is to assist analysts in identifying attack stages from a community perspective, enabling a quick understanding of the purpose of a subgraph in the provenance graph (e.g., file compilation and uploading). HERCULE and WATSON (subgraph-level) differ in granularity from SPARSE (event-level). Hence, we do not compare our work with them.
\par
Here, we compare the performance of \sn{} with 3 state-of-the-art approaches: SLEUTH~\cite{hossain2017sleuth}, NODOZE~\cite{hassan2019nodoze}, and DEPIMPACT~\cite{fang2022depimpact}, which are more relevant in terms of methodology (anomaly-score based) and granularity (event-level) for our evaluation. SLEUTH defines TTP-like rules with the added constraint that these rules only fire when certain confidentiality or integrity conditions are satisfied according to a tag-based information flow propagation. NODOZE measures the rarity of different events in the environment and based on this assigns anomaly scores to each event in the dependency graph. We use logs that only contain normal behavior (captured outside of attack periods) as execute profiles (i.e., statistics of events) to satisfy NODOZE. DEPIMPACT assigns anomaly scores to edges based on a number of characteristics (including time, data flow amount, and node access), and then aggregates the scores to determine the entry points through a propagation algorithm. DEPIMPACT then takes as output the overlap events of the forward graph of the entry point and the backward graph of the alert point. Finally, we also perform an ablation experiment on \sn{} to evaluate the output of different phases.

\par
Table~\ref{tab:diff_tech_per} shows the performance of attack investigation for different techniques in all cases. Lower FP/FPR indicate a better ability to filter irrelevant edges and lower FN/FNR indicate a better ability to retain critical edges. The results show that \sn{} (SSG + PCA) performs the best. On average, the critical component graph generated by \sn{} ($\sim$ 113 edges) is 8849 $\times $ smaller than the original dependency graph ($\sim$ 1,000,000 edges), 22 $\times$ smaller than the second-best result (i.e., DEPIMAPCT with $\sim$ 2,487 edges). \sn{} demonstrates the best capability in filtering irrelevant edges while preserving the attack sequences (FP = 99, FPR = 0.043*$10^{-2}$),  25 $\times$ more effective than DEPIMPACT (FP = 2,473, FPR = 1.087*$10^{-2}$). Moreover, \sn{} does not miss any critical edges (i.e., FN = 0, FNR = 0). Note that Column ``SSGC'' and Column ``SFP'' denote the \textit{suspicious semantic graph construction} and \textit{suspicious flow path} of the intermediate product of \sn{}, respectively.    
\par
SLEUTH investigates attack scenarios by defining confidentiality and integrity labels for label propagation. However, SLEUTH cannot ensure to cover all the attack-related edges and therefore performed the worst in FNR (FN = 5.73, FPR = 42.54*$10^{-2}$), resulting in ineffective support for attack investigation. NODOZE performed better than SLEUTH in including critical edges but worse in FPR (FP = 16,682, FPR = 7.329*$10^{-2}$; FN = 2, FNR = 14.85*$10^{-2}$), resulting in the ineffective reduction of investigation cost for analysts. The reason is that the performance of NODOZE relies on whether the execution profile comprehensively covers all benign behaviors but ignores information in the form of streams. DEPIMPACT heuristically selects 3 entry points (one each for files, processes, and sockets), which amplifies the attack surface. DEPIMPACT directly takes the intersection between the forward graph of the entry points and the backward graph of the alert point as output, which introduces massive attack-irrelevant events (FP = 2,473, FPR = 1.087*$10^{-2}$). Regarding the reduction results, \sn{} exhibits the best performance (FP = 99, FPR = 0.043*$10^{-2}$), we attribute the good performance of our system to (1)  \sn{} employs insight of semantic transfer to construct a suspicious semantic graph related to the POI event that inherently filters out a significant number of irrelevant events (FP = 403, FPR = 0.177*$10^{-2}$), and (2) \sn{} evaluates the relevance of context to POI events at the path-level rather than in isolation.
\par
In summary, we believe that \sn{} can satisfy the requirement of low false positives in practice.
\begin{table}[]
\footnotesize
\centering
\caption{Comparison of data consumption rate and data generation rate.}
\label{tab:events_consume}
\resizebox{0.45\textwidth}{!}{%
\begin{tabular}{crr}
\hline
\textbf{\begin{tabular}[c]{@{}c@{}}Attack\\ Case\end{tabular}} & \multicolumn{1}{c}{\textbf{\begin{tabular}[c]{@{}c@{}}Generation Rate\\ (event num/pers)\end{tabular}}} & \multicolumn{1}{c}{\textbf{\begin{tabular}[c]{@{}c@{}}Consumption Rate\\ (event num/pers)\end{tabular}}} \\ \hline
Wget Executable                                                & 3,018                                                                                                   & 42,175                                                                                                   \\
Illegal Storage                                                & 3,273                                                                                                   & 45,572                                                                                                   \\
Illegal Storage2                                               & 3,156                                                                                                   & 34,727                                                                                                   \\
Hide File                                                      & 4,039                                                                                                   & 31,155                                                                                                   \\
Steal Information                                              & 4,271                                                                                                   & 40,719                                                                                                   \\
Backdoor Dowanload                                             & 3,380                                                                                                   & 39,451                                                                                                   \\
Annoying Server User                                           & 3,153                                                                                                   & 42,155                                                                                                   \\
Shellshock                                                     & 3,741                                                                                                   & 42,742                                                                                                   \\
Dataleak                                                       & 3,692                                                                                                   & 42,899                                                                                                   \\
VPN Filter                                                     & 3,560                                                                                                   & 39,611                                                                                                   \\
Five Dir Case 1                                                & 917                                                                                                     & 40,358                                                                                                   \\
Five Dir Case 3                                                & 1,161                                                                                                   & 44,795                                                                                                   \\
Theia Case 1                                                   & 1,298                                                                                                   & 32,097                                                                                                   \\
Theia Case 3                                                   & 1,215                                                                                                   & 34,856                                                                                                   \\
Theia Case 5                                                   & 966                                                                                                     & 41,303                                                                                                   \\ \hline
\textbf{Avg}                                                   & 2,722                                                                                                   & 39,641                                                                                                    \\ \hline
\end{tabular}%
}
\end{table}

\begin{table*}[]
\centering
\footnotesize
\caption{\centering Overhead performance of each component of \sn{} and baseline approach.}
\label{tab:overhead}
\resizebox{\textwidth}{!}{%
\begin{tabular}{c|rrr|rrrr|rrrr}
\hline
\multirow{2}{*}{\textbf{Attack}} & \multicolumn{3}{c|}{\textbf{SSGC}}                                                                                                                                                                                                                       & \multicolumn{4}{c|}{\textbf{PCA}}                                                                                                                                                                                                                                                                                                        & \multicolumn{4}{c}{\textbf{DEPIMPACT}}                                                                                                                                                                                                                                                                                                    \\
                                 & \multicolumn{1}{c}{\textbf{\begin{tabular}[c]{@{}c@{}}Mem.\\ (MB)\end{tabular}}} & \multicolumn{1}{c}{\textbf{\begin{tabular}[c]{@{}c@{}}CPU \\ (\%)\end{tabular}}} & \multicolumn{1}{c|}{\textbf{\begin{tabular}[c]{@{}c@{}}Disk \\ (MB)\end{tabular}}} & \multicolumn{1}{c}{\textbf{\begin{tabular}[c]{@{}c@{}}Ti. \\ (s)\end{tabular}}} & \multicolumn{1}{c}{\textbf{\begin{tabular}[c]{@{}c@{}}Mem. \\ (MB)\end{tabular}}} & \multicolumn{1}{c}{\textbf{\begin{tabular}[c]{@{}c@{}}CPU \\ (\%)\end{tabular}}} & \multicolumn{1}{c|}{\textbf{\begin{tabular}[c]{@{}c@{}}\# \\ SFP\end{tabular}}} & \multicolumn{1}{c}{\textbf{\begin{tabular}[c]{@{}c@{}}Ti. \\ (s)\end{tabular}}} & \multicolumn{1}{c}{\textbf{\begin{tabular}[c]{@{}c@{}}Mem.\\ (MB)\end{tabular}}} & \multicolumn{1}{c}{\textbf{\begin{tabular}[c]{@{}c@{}}CPU \\ (\%)\end{tabular}}} & \multicolumn{1}{c}{\textbf{\begin{tabular}[c]{@{}c@{}}Disk \\ (MB)\end{tabular}}} \\ \hline
Wget Executable                  & 29.75                                                                            & 3.14                                                                             & 0.072                                                                              & 0.09                                                                            & 108.20                                                                            & 4.97                                                                             & 5                                                                               & 4.45                                                                            & 125.72                                                                           & 7.62                                                                             & 0.79                                                                              \\
Illegal Storage                  & 29.92                                                                            & 3.32                                                                             & 7.15                                                                               & 0.32                                                                            & 124.97                                                                            & 4.36                                                                             & 85                                                                              & 285.10                                                                          & 178.89                                                                           & 7.46                                                                             & 79.21                                                                             \\
Illegal Storage2                 & 31.81                                                                            & 3.35                                                                             & 69.21                                                                              & 2.38                                                                            & 125.16                                                                            & 4.27                                                                             & 522                                                                             & 985.55                                                                          & 304.51                                                                           & 7.29                                                                             & 169.21                                                                            \\
Hide File                        & 34.82                                                                            & 3.59                                                                             & 101.12                                                                             & 6.22                                                                            & 147.34                                                                            & 4.94                                                                             & 833                                                                             & 19,814.25                                                                       & 592.82                                                                           & 7.51                                                                             & 401.12                                                                            \\
Steal Information                & 36.55                                                                            & 3.61                                                                             & 109.17                                                                             & 6.66                                                                            & 155.60                                                                            & 4.90                                                                             & 868                                                                             & 12,701.39                                                                       & 596.26                                                                           & 7.82                                                                             & 409.17                                                                            \\
Backdoor Dowanload               & 28.01                                                                            & 3.27                                                                             & 2.21                                                                               & 0.10                                                                            & 125.16                                                                            & 4.49                                                                             & 20                                                                              & 33.68                                                                           & 199.17                                                                           & 7.42                                                                             & 61.13                                                                             \\
Annoying Server User             & 26.62                                                                            & 3.40                                                                             & 0.011                                                                              & 0.33                                                                            & 119.41                                                                            & 4.41                                                                             & 6                                                                               & 0.23                                                                            & 127.81                                                                           & 7.76                                                                             & 0.70                                                                              \\
Shellshock                       & 29.74                                                                            & 3.39                                                                             & 8.94                                                                               & 0.02                                                                            & 137.75                                                                            & 4.13                                                                             & 40                                                                              & 2.17                                                                            & 138.98                                                                           & 7.57                                                                             & 154.00                                                                            \\
Dataleak                         & 30.39                                                                            & 3.40                                                                             & 0.82                                                                               & 0.04                                                                            & 132.86                                                                            & 4.29                                                                             & 17                                                                              & 2.24                                                                            & 130.65                                                                           & 7.22                                                                             & 97.50                                                                             \\
VPN Filter                       & 31.01                                                                            & 3.26                                                                             & 0.82                                                                               & 0.06                                                                            & 146.08                                                                            & 4.27                                                                             & 37                                                                              & 15.46                                                                           & 135.49                                                                           & 7.43                                                                             & 537.95                                                                            \\
Five Dir Case 1                  & 28.83                                                                            & 3.29                                                                             & 7.15                                                                               & 0.04                                                                            & 124.28                                                                            & 4.56                                                                             & 11                                                                              & 2.08                                                                            & 124.12                                                                           & 7.67                                                                             & 0.82                                                                              \\
Five Dir Case 3                  & 30.05                                                                            & 3.50                                                                             & 37.71                                                                              & 0.11                                                                            & 135.04                                                                            & 4.29                                                                             & 22                                                                              & 1,674.90                                                                        & 189.08                                                                           & 7.72                                                                             & 88.65                                                                             \\
Theia Case 1                     & 31.97                                                                            & 3.43                                                                             & 59.50                                                                              & 2.34                                                                            & 137.35                                                                            & 4.15                                                                             & 374                                                                             & 24,761.90                                                                       & 844.28                                                                           & 7.64                                                                             & 479.20                                                                            \\
Theia Case 3                     & 35.08                                                                            & 3.62                                                                             & 68.45                                                                              & 4.85                                                                            & 143.90                                                                            & 4.80                                                                             & 598                                                                             & 36,682.87                                                                       & 885.83                                                                           & 7.79                                                                             & 592.23                                                                            \\
Theia Case 5                     & 29.12                                                                            & 3.30                                                                             & 0.13                                                                               & 0.19                                                                            & 129.21                                                                            & 4.67                                                                             & 15                                                                              & 4.91                                                                            & 138.60                                                                           & 7.15                                                                             & 1.41                                                                              \\ \hline
\textbf{Avg}                     & 30.91                                                                            & 3.39                                                                             & 21.03                                                                              & 1.58                                                                            & 132.82                                                                            & 4.50                                                                             & 230                                                                             & 6464.75                                                                         & 320.81                                                                           & 7.54                                                                             & 204.87                                                                            \\ \hline
\end{tabular}%
}
\begin{tablenotes}
    \centering
    \footnotesize
    \item $\dagger $ \textbf{SSGC}:Suspicious Semantic Graph Construction.  \textbf{PCA}:Path-level Contextual Analysis. \textbf{SFP}:Suspicious Flow Path. 
\end{tablenotes}
\end{table*}

\par
\subsection{RQ2 : How efficient is \sn{} in attack investigation ?}
\label{sec:5.3_efficient}
In this section, we evaluate the efficiency of \sn{} when deployed in a real scenario. First, we evaluate \sn{} on the real-time performance by comparing data generation rate and data consumption rate. Table~\ref{tab:events_consume} shows that \sn{} can consume events 15 $\times$ faster than the events generation rate of the host on average, which shows the real-time of \sn{} is feasible. Then we evaluate the overheads of each component of \sn{} on time, memory, CPU, and disk, as shown in Table~\ref{tab:overhead}. In addition, we conduct comparative experiments with the baseline method DEPIMPACT.
\par
\textbf{Time and Memory.} \sn{} can perform path-level contextual analysis of the suspicious semantic graph and construct the critical component graph in 2s on average, which is 4091 $\times$ faster than DEPIMPACT ($\sim$ 6,464s). DEPIMPACT uses an impact propagation algorithm to identify entry points. However, when the number of edges in the backtracking graph is high (e.g., case ``Hide File''), it takes an extremely long time to reach global convergence (19,814s). 
In terms of memory overhead, three causes are making \sn{} smaller than DEPIMAPCT: (1) \sn{} only stores suspicious nodes in memory when processing streaming logs, so the memory overhead grows extremely slowly and can be seen as constant in scale (i.e., 30MB). (2) The suspicious semantic graphs ($\sim$ 403 edges) read in by \sn{} are much smaller than the dependency graphs ($\sim$ one million edges) read in by DEPIMPACT. (3) The suspicious flow path extraction algorithm applied in PCA only traverses the graph structure once (a complexity of $O(E)$) and generates a smaller number of SFPs (230 on average). For these reasons, the memory overhead of PCA (132.82 MB) is 2.4 $\times$ smaller than that of DEPIMAPCT (320.81 MB). Finally, it is important to emphasize that \sn{} is real-time for suspicious semantic graph construction (SSGC), so we do not perform a time overhead evaluation for this component.
\par
\textbf{CPU and Disk.} \sn{} stores the relevant event table (RET) in the database when processing streaming logs. To evaluate the overhead of \sn{} on the hard disk, we also perform the relevant experiments. As shown in Table~\ref{tab:overhead}, \sn{} requires only 21.03 MB of disk space, which is 9 $\times$ smaller than the original logs (204.87 MB). This is because \sn{} only retains events of suspicious semantic relevance, whereas other techniques~\cite{tang2018nodemerge, hassan2019nodoze, fang2022depimpact, ma2016protracer, liu2018priortracker} require the whole logs, which significantly increases overhead on the disk. In addition, the CPU overhead of \sn{} is 4.5\%, due to the simple yet intuitive and effective algorithm.
\par
In summary, we believe that \sn{} outperforms the latest work DEPIMPACT in all aspects of overhead and can satisfy the requirements of low overhead and latency in application scenarios.

\subsection{RQ3 : How sensitive is \sn{} in parameter selection?}

\begin{figure*}[h!t]
\centering
\setlength{\abovecaptionskip}{-0.1cm}
\subfigure[Hyperparameter matrix on $FP$]{\includegraphics[width=8cm]{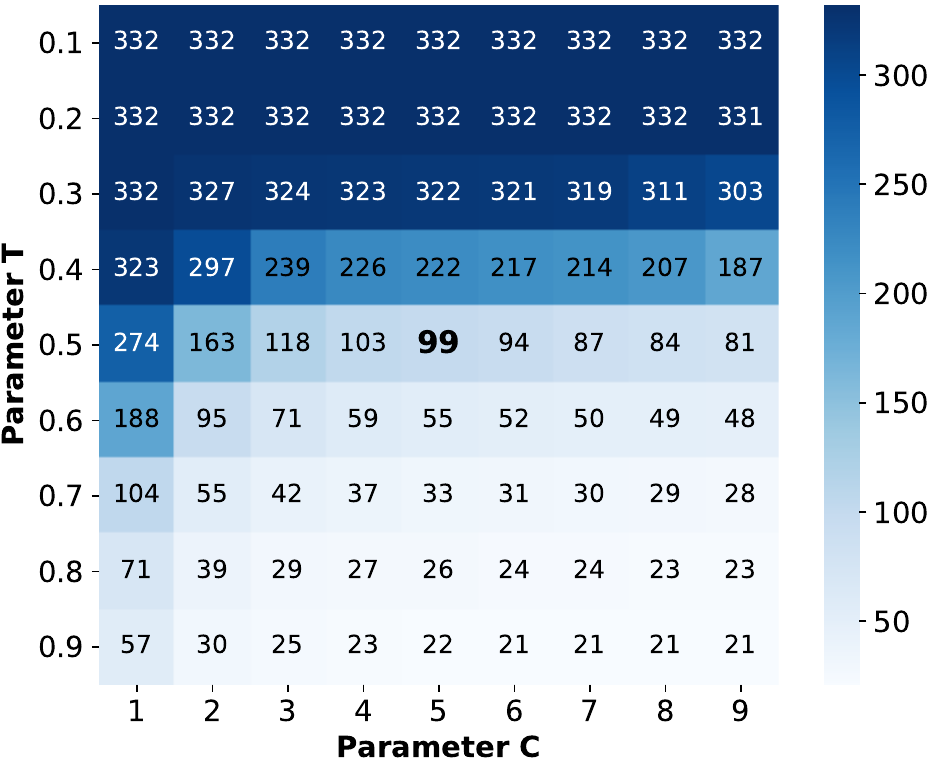}\label{fig:conf_matrix_FP}}
\subfigure[Hyperparameter matrix on $FN$]{\includegraphics[width=8cm]{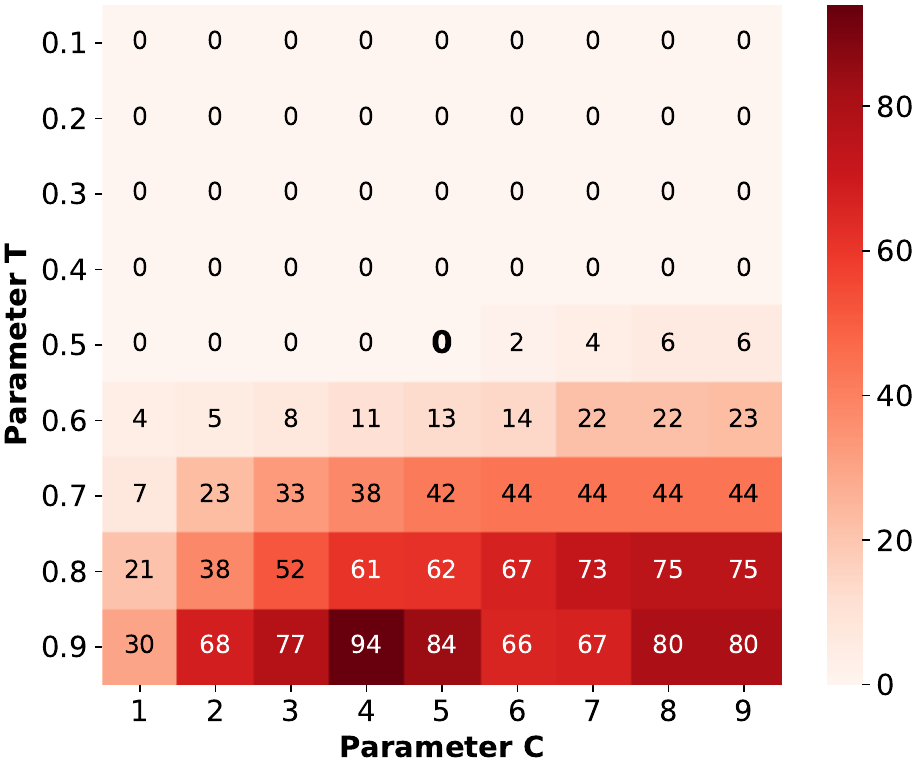}\label{fig:conf_matrix_FN}}
\caption{Hyperparameter Matrices for System Performance with Different Parameters.}
\label{fig:para}
\vspace{-0.2cm}
\end{figure*}

\par
As shown in Section~\ref{sec:4.6_pathScroing}, there are two super parameters in \sn{} that need to be set: (1) the $C$ affects the expansion factor, (2) the $T$ filters the paths. As shown in Equation~\ref{euqation3}, the larger $C$, the smaller the expansion factor $\alpha$,  the lower the path score, and the fewer events will be retained by \sn{}. Parameter $T$, on the other hand, indicates the severity of \sn{} for path selection; the larger $T$ is, the fewer events will be retained by \sn{}.
\par
We demonstrate the sensitivity of \sn{} in parameter selection by testing all combinations of the hyper-parameters $C$ and $T$ through grid search methodology. Specifically, we set the minimum value of $C$ to be 1, the maximum value to be 9, and the step size to be 1; set the minimum value of $T$ to be 0.1, the maximum value to be 0.9, and the step size to be 0.1; and test the performance of \sn{} on the metrics FP/FN as shown in Figure~\ref{fig:para}. As parameter $C$ increases, \sn{} reduces the path score and retains fewer events, hence the FP decreases. At the same time, \sn{} misses some critical events, causing FN to rise. As the threshold $T$ increases, \sn{} blocks more paths and preserves fewer events, so FP falls while FN rises. As shown in Figure~\ref{fig:conf_matrix_FP}, FP decreases gradually from the upper left to the lower right; as shown in Figure~\ref{fig:conf_matrix_FN}, FN increases gradually from the upper left to the lower right. The effects of these parameter changes on the \sn{} are in line with our predictions.
\par
Obviously, FP and FN are two evaluation metrics that we both want to be as low as possible, but there is a trade-off between their performance for a system (i.e., a rise in one leads to a fall in the other). Finally, we choose $C$ = 5 and $T$ = 0.5 as the default parameters for \sn{}. Of course, the manufacturer can adapt these parameters to the specific scenario.

\section{Discussion}
\label{sec:6_discussion}

\textbf{Cooperation with existing techniques.} There is a requirement for defenders to be able to detect and handle real-world attacks in real time. As an attack investigation system, \sn{} can be combined with a variety of existing techniques to meet this goal. By working with intrusion detection systems~\cite{milajerdi2019holmes, zhu2023aptshield, xiong2020conan, kim2010intrusion} that can provide real-time alerts and defenses, \sn{} is able to investigate alerts for relevant events and provide a brief critical component graph to analysts. By working with compression systems~\cite{xu2016cpr, tang2018nodemerge, hossain2018dependence, michael2020forensic, zhu2021gs-ss} that can reduce redundant information, \sn{} is able to reduce memory and disk overheads, enabling years of relevant data storage.  By working with analysis systems~\cite{hassan2020rapsheet, hassan2020swift, hassan2019nodoze} that automatically determine the authenticity of alarms, \sn{} is able to provide streamlined but sufficient relevant events to support the triage of alarms.
\par
\textbf{Evasion Attacks.} Existing investigation techniques, such as DEPIMAPCT, utilize weight computation and score-propagation techniques to identify attack entry points. However, this insight of independently calculating the weights of events does not fit the situation where information flows between entities. As a result, an attacker can inject a payload by writing multiple times and thus evade tracking. In contrast, \sn{} mitigates the impact of this by performing contextual analysis in path-level to synthesize the relevance between a path and an alert. An attacker may evade investigation by going the long way around (i.e., repeating nonsensical behavior) as in the attack case ``Hide File'' where the attacker changes the file name multiple times. As shown in Section~\ref{sec:4.6_pathScroing}, \sn{} is able to mitigate this problem by performing relative score calculation and event score inflation mechanisms.
\par
\textbf{Limitation.} To implement attack investigation, \sn{} relies on alerts initiated by Endpoint Detection and Response (EDR) placed on the host. \sn{} cannot perform attack investigation if the detection system fails to launch alerts (identify suspicious behavior). 
Recent approaches~\cite{wang2019heterogeneous, han2020unicorn} propose solutions to improve the detection of anomalous system activity, and \sn{} can work with these approaches to provide better defenses. If the detection system initiates false positives frequently, \sn{} can only identify relevant events but cannot filter these false alters. But \sn{} can work with alarm triage techniques~\cite{hassan2020rapsheet, hassan2019nodoze, hossain2018dependence} to help them filter false alarms by providing a streamlined critical component graph. 
In addition, as shown in Section~\ref{sec:4.3.3_storage}, \sn{} needs to maintain a suspicious entity list  in memory and a  related event table on disk. As the runtime grows, there is redundant information in the related event table, such as $Event\ 10$ being stored four times in Figure~\ref{fig:semanTransExample}.  By working with existing compression systems~\cite{xu2016cpr, tang2018nodemerge, hossain2018dependence, michael2020forensic, zhu2021gs-ss}, \sn{} can effectively mitigate this situation and enable long deployment runs.
\section{Related Work}
\label{sec:7_relatedwork}
The analysts need to perform threat alert validation and post-mortem analysis of incidents. Currently, while auditing is by no means the only form of forensic investigation, it is telling that 75\% of incident response specialists consider logs to be the most valuable form of investigation artifact~\cite{carbon-black}. Several studies have focused on implementing efficient log collection, such as kellect~\cite{chen2022kellect} for Windows and e-bpf~\cite{byrnes2020modern, wang2022design} for Linux, which is a pre-task for attack investigation. And in terms of methodologies for attack investigation using logs, they can be categorized into three categories: label propagation-based, anomaly score-based, and machine learning-based.
\par
\textbf{Label Propagation-based.} Labels are given to nodes and are propagated to other nodes by system calls. When an alert arises, the analyst can easily retrace the events associated with the alert based on the label. Milajerdi et al. propose HOLMES~\cite{milajerdi2019holmes}, which mitigates the dependency explosion problem by requiring the aggregation of more labels to raise the detection threshold. RapSheet~\cite{hassan2020rapsheet} makes use of the tactical provenance graph (TPG), which instead of encoding low-level system event dependencies, reasons about the causal relationships between threat alerts. RapSheet proposes a threat scoring scheme that evaluates the severity of each alert based on TPGs to enable effective investigation of alerts. Zhong et al.~\cite{zhong2016automate}, on the other hand, mine the analysts' security operation traces to learn label-propagation rules, and then use these rules to identify relevant paths automatically when an alert occurs. CONAN~\cite{xiong2020conan} iteratively performs malicious behavior determination with label passing and aggregation through data provided by ETW, enabling real-time detection and investigation. These label-based approaches rely on heuristic rules that cannot handle all types of attacks and have a high level of false negatives in attack investigation.
\par
\textbf{Anomaly Score-based.} The essential view of these methods is to quantify the suspiciousness of edges between pairs of nodes. Pei et al.'s HERCULE~\cite{pei2016hercule} system correlates multi-source heterogeneous logs to construct a multi-dimensional weighted graph and uses the unsupervised community detection algorithm Louvain~\cite{blondel2008louvain} to discover attack-related paths from it. NODOZE~\cite{hassan2019nodoze} and PRIOTRACKER~\cite{liu2018priortracker}, on the other hand, perform statistics on historical data and assign anomaly scores to events in the dependency graph. The score propagation algorithm is then used to find suspicious events. Both of these methods rely on statistics of historical data and cannot be applied in complex and variable generative environments. DEPIMPACT~\cite{fang2022depimpact} calculates dependency weights globally based on multiple features (including time, data traffic amount, and node access) and then aggregates the weights to nodes to determine suspicious points of intrusion. The overlapping parts between the forward graph of the entry point and the backward graph of the alarm point are then considered attack-related events. These score-based methods can cover all critical events. However, there is no restriction on the variation of scores to solve the dependency explosion problem, thus leading to higher false positives. In addition, these methods require reading in all the logs to build the dependency graph, which is very expensive in terms of hard disk and memory.
\par
\textbf{Machine Learning-based.} Some techniques use machine learning methods to learn contextual and structural information from dependency graphs to identify the most relevant abnormal events to alert. ATLAS~\cite{alsaheel2021atlas} uses a novel combination of causal analysis, natural language processing, and machine learning to construct sequence-based models as a way to establish critical patterns of attack and non-attack behavior in the dependency graph. On the other hand, DEPCOMM~\cite{xu2022depcomm} proposes a novel graph summarization method by dividing the large graph into process-centric subgraphs. DEPCOMM then extracts summaries from each subgraph, enabling the generation of summary graphs from dependency graphs, thereby reducing the difficulty of investigation for analysts. WATSON~\cite{zeng2021watson} automatically abstracts and clusters high-level system behavioral features from low-level audit events. WATSON performs a Depth-First Search on each object to summarise system behavior and then uses machine learning to infer the semantics of each audit event based on its context. Finally, behaviors with similar semantics are aggregated in the embedding space to identify similar events to the alert in order to investigate the attack. These machine learning-based approaches suffer from inadequate training samples, poor generalization capabilities, and high computational costs.
\par
In contrast to previous work, \sn{} employs a hybrid method in the specific domain of causality tracking. \sn{} first uses \textit{suspicious semantic delivery rule} to construct suspicious semantic graph. Then \sn{} uses \textit{path-level contextual analysis} to extract a streamlined critical component graph.
\section{Conclusion}
\label{sec:8_conclusion}
We propose \sn{}, a system that processes streaming logs and outputs critical events (attack-related events) according to alert in real-time. Specifically, \sn{} constructs a suspicious semantic graph related to the POI event by \textit{suspicious semantic transfer rule and storage strategy}. Then \sn{} uses a \textit{suspicious flow path extraction algorithm} to extract all reachable flow paths from the suspicious semantic graph. Finally, \sn{} uses \textit{path-level contextual analysis} to score all paths and filters irrelevant events to obtain the final critical component graph. Our evaluation of real attacks demonstrates that \sn{} achieves low false positives (FP = 99), low overhead (30MB for memory and 21.03MB for hard disk), and low latency (1.58s for attack investigation).

\bibliographystyle{IEEEtran}
\bibliography{main}

\clearpage
\appendix
\setcounter{equation}{0}
\pagestyle{empty}
\renewcommand{\thesection}{Appendix \arabic{section}}
\setcounter{section}{0} 
\section{Appendix}
\label{appendix}
\subsection{Attack Cases}
\label{sec:appendix_attacksample}
In this section, we show the ground truth for the 10 attack cases used for evaluation in Section~\ref{sec:5.1.2_groundtruth}. As shown in Figure~\ref{fig:7_attacks} and Figure~\ref{fig:3_attacks}, for entities, we use rectangles ($\langle ProcessName,\ Prcoess ID \rangle$) for processes, ellipses ($\langle FileName \rangle$) for files, and diamonds ($\langle SrcIP\ :\ SrcPort \rangle$) for sockets. For events, we use solid lines with arrows, where the arrows indicate the flow of information. In addition, the number on the solid line indicates the relative time of the event. The solid red line indicates the POI event that triggered the alarm.
\par
\subsubsection{Attacks Based on Commonly Used Exploits}\label{sec:appendix_common-7}
\begin{figure*}[]
\setlength{\abovecaptionskip}{0.cm}
\caption{The ground truth of 7 single-host attack cases.}
\label{fig:7_attacks}
\centering
\subfigure[Wget Executable]{\includegraphics[width=0.32\hsize, height=0.3\hsize]{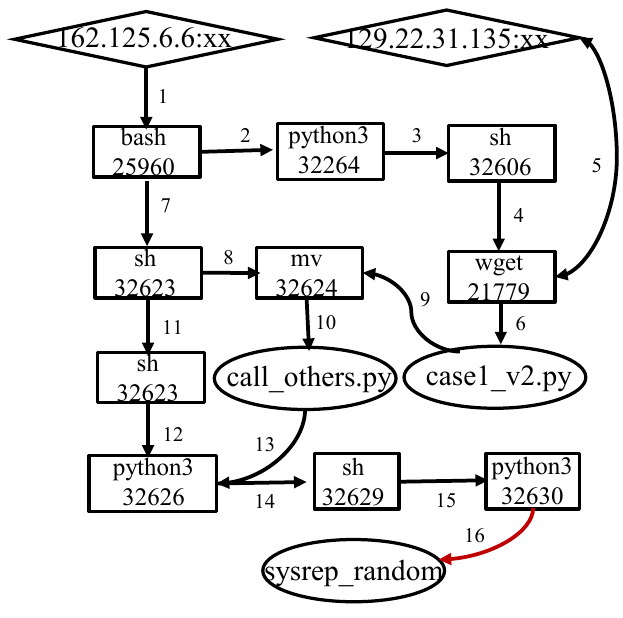}\label{fig:case1}}
\subfigure[Illegal Storage]{\includegraphics[width=0.32\hsize, height=0.3\hsize]{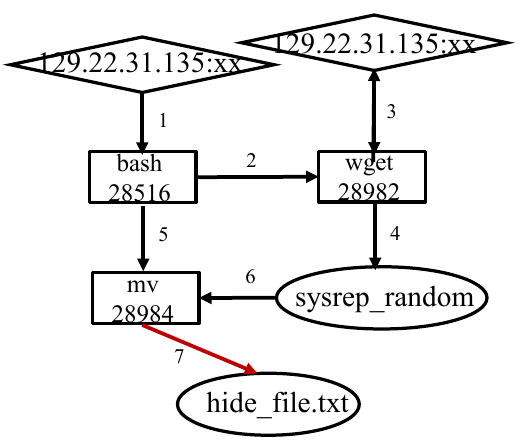}\label{fig:case2}}
\subfigure[Illegal Storage 2]{\includegraphics[width=0.32\hsize, height=0.3\hsize]{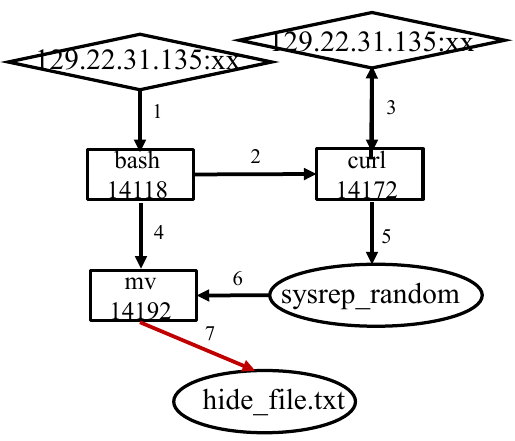}\label{fig:case3}}
\subfigure[Hide File]{\includegraphics[width=0.32\hsize, height=0.3\hsize]{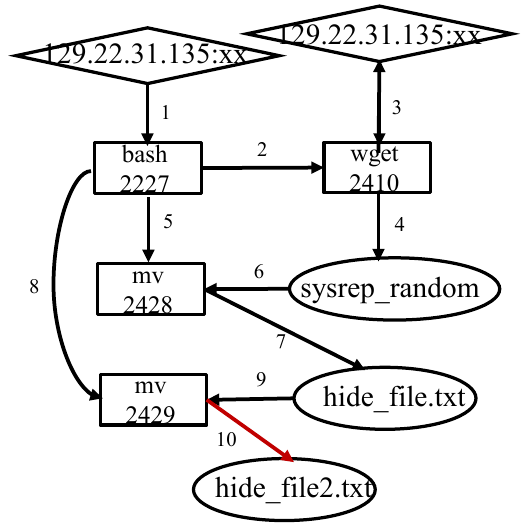}\label{fig:case4}}
\subfigure[Steal Information]{\includegraphics[width=0.32\hsize, height=0.3\hsize]{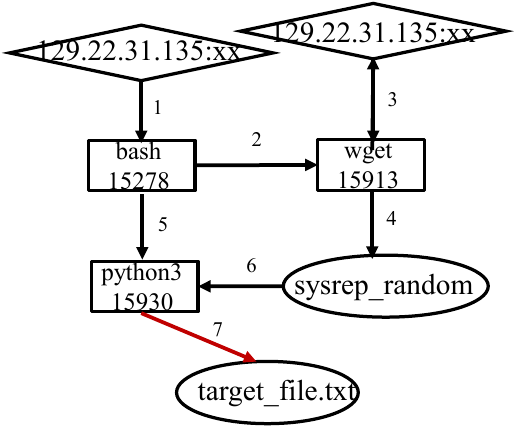}\label{fig:case5}}
\subfigure[Backdoor Download]{\includegraphics[width=0.32\hsize, height=0.3\hsize]{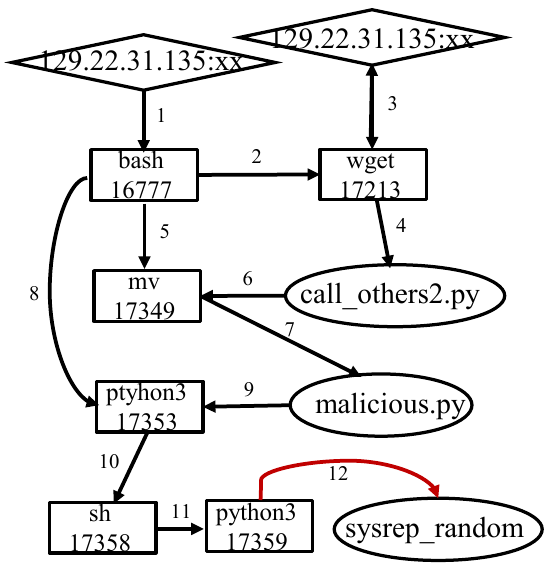}\label{fig:case6}}

\subfigure[Annoying Server User]{\includegraphics[width=0.32\hsize, height=0.3\hsize]{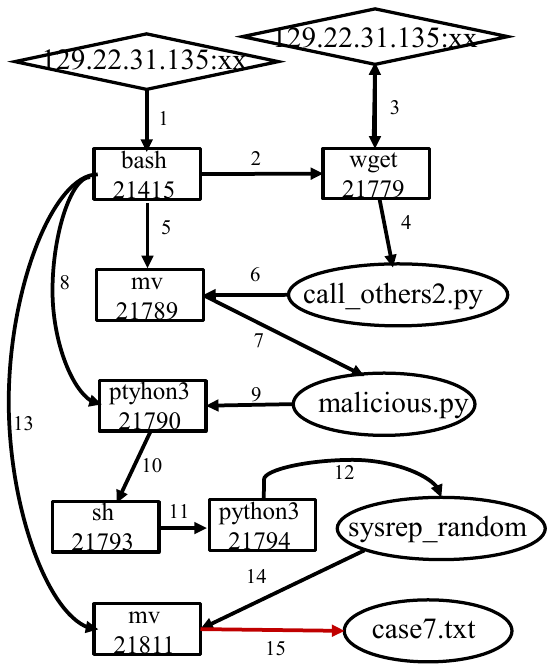}\label{fig:case7}}
\end{figure*}

\ \\
These 7 attacks are applied in the evaluations of previous works~\cite{ma2016protracer, xu2016cpr, fang2022depimpact, kwon2018mci}, and consisted of the following scenarios:
\begin{itemize}
\item{\textit{Wget Executable~\cite{xu2016cpr}}: Unsecured servers expose a vulnerability, allowing unauthorized users to fetch executable Python scripts through wget and execute them, as shown in Figure~\ref{fig:7_attacks}(a).}
\item{\textit{Illegal Storage~\cite{ma2016protracer}}: Leveraging wget, a server administrator retrieves suspicious files and deposits them into a user's home directory, as shown in Figure~\ref{fig:7_attacks}(b).}
\item{\textit{Illegal Storage 2~\cite{ma2016protracer}}: Leveraging curl, a server administrator retrieves suspicious files and deposits them into a user's home directory, as shown in Figure~\ref{fig:7_attacks}(c).}
\item{\textit{Hide File~\cite{kwon2018mci}}: With the intention of concealing a malicious file among user's normal files, the attacker downloads a script and obfuscates it by altering the filename and location, as shown in Figure~\ref{fig:7_attacks}(d).}
\item{\textit{Steal Information~\cite{ma2016protracer}}: The attacker steals user's sensitive data and stores it in a covert file, avoiding detection, as shown in Figure~\ref{fig:7_attacks}(e).}
\item{\textit{Backdoor Download~\cite{ma2016protracer}}: A malicious insider establishes a connection to a rogue server using the ping command. Subsequently, the insider downloads a concealed backdoor script and hides the script, as shown in Figure~\ref{fig:7_attacks}(f).}
\item{\textit{Annoying Server User~\cite{kwon2018mci}}: A malicious user, gaining access to other users' home directories, injects superfluous data into their files, as shown in Figure~\ref{fig:7_attacks}(g).}
\end{itemize}

\subsubsection{Multi-host Intrusive Attacks}\label{sec:appendix_multi-3}
\par
\begin{figure*}[]
\setlength{\abovecaptionskip}{0.cm}
\caption{The ground truth of 3 multi-host attack cases.}
\label{fig:3_attacks}
\centering
\subfigure[Shesllshock]{\includegraphics[width=0.8\hsize, height=0.33\hsize]{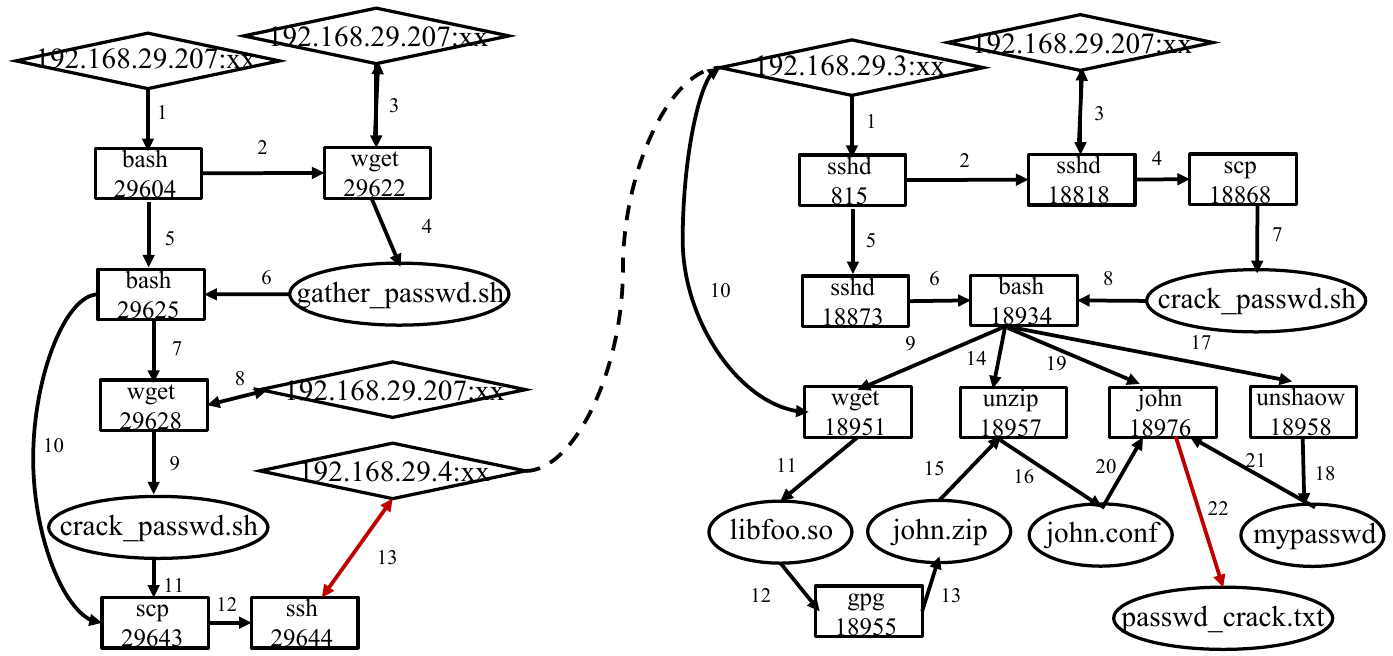}\label{fig:case8}}

\subfigure[Dataleak]{\includegraphics[width=0.8\hsize, height=0.33\hsize]{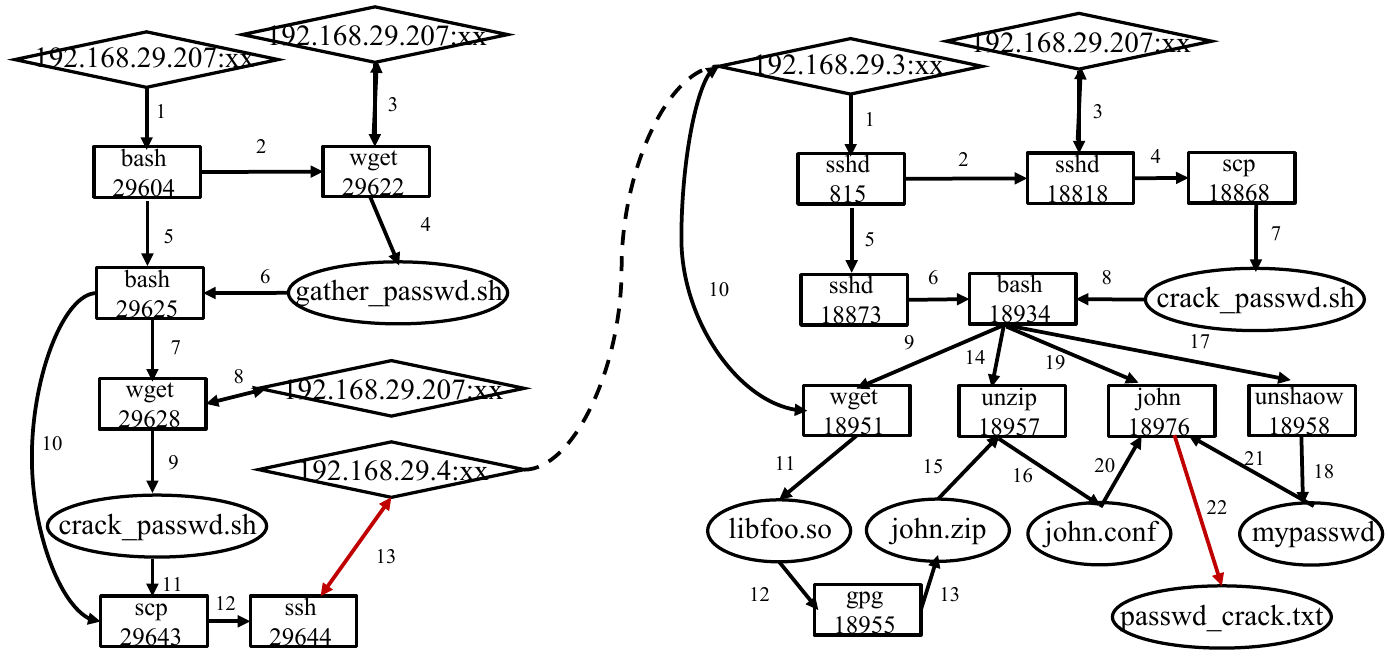}\label{fig:case9}}

\subfigure[VPN Filter]{\includegraphics[width=0.8\hsize, height=0.33\hsize]{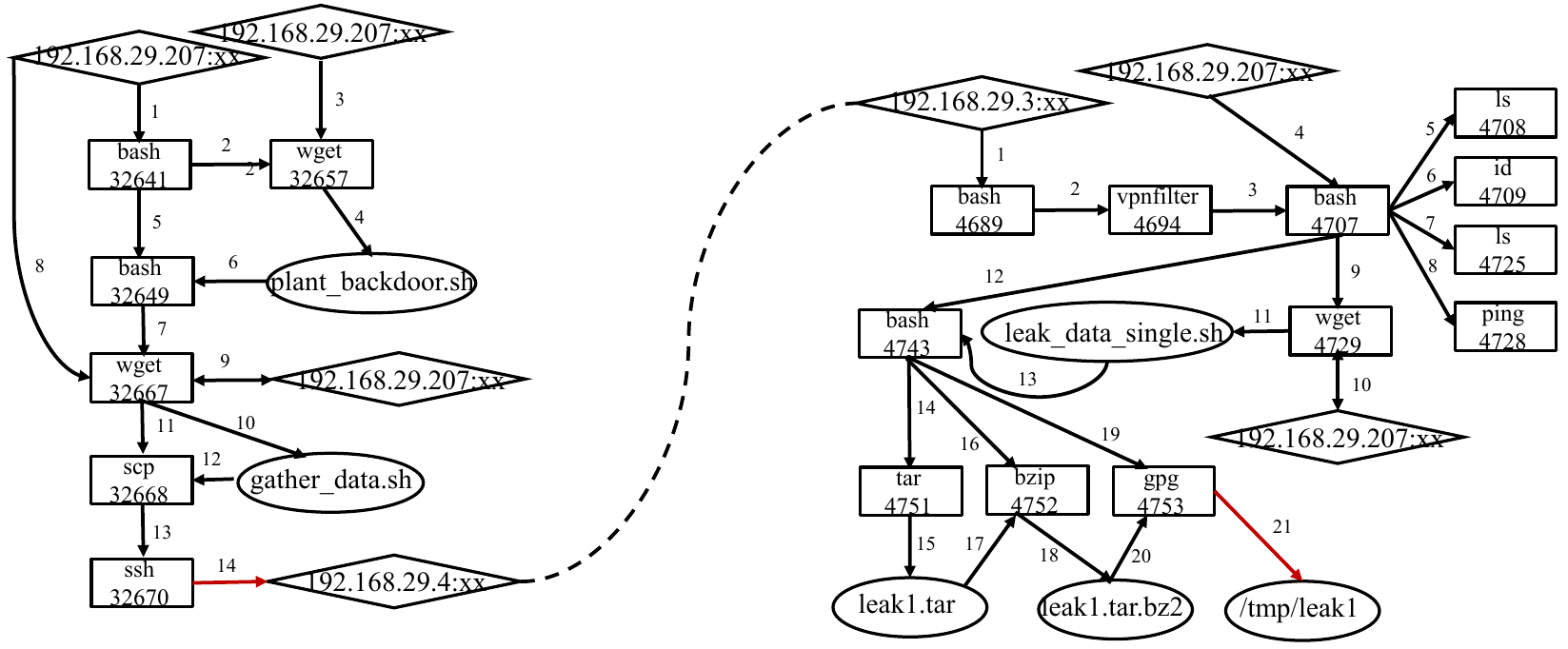}\label{fig:case10}}

\end{figure*}

In \textbf{Attack 1}, known as \textbf{Shellshock Penetration}, the attacker, following the initial exploit on Host 1, establishes a connection to cloud services (e.g., Dropbox, Twitter). Here, an image containing the C2 server's IP address encoded in the EXIF metadata is downloaded. This tactic, reminiscent of advanced persistent threat (APT) attacks~\cite{vpnfilter, Ebay}, aims to evade network-based detection systems relying on DNS blacklisting. Leveraging the obtained IP address, the attacker proceeds to download malware from the C2 server to Host 1. Upon execution of the script, an examination of the ssh configuration file ensues, revealing reachable hosts in the network, including Host 2, Host 3, and Host 4. Subsequently, the malware fetches another script from the C2 server and disseminates it to the identified hosts, extracting passwords in the process, as shown in Figure~\ref{fig:3_attacks}(a).
\par
In \textbf{Attack 2}, known as \textbf{Data Leakage}, the attacker, post-reconnaissance, acquires another malware, \textit{leak\_data.sh}, from the C2 server, distributing it to Host 2. This malware scans for concealed files and files containing sensitive strings, compressing them into a tarball named \textit{leak.tar.bz2}. The compressed tarball is then transmitted back to Host 1, where it undergoes encryption before being uploaded to the internet, as shown in Figure~\ref{fig:3_attacks}(b).
\par
In \textbf{Attack 3}, known as \textbf{VPN Filter}~\cite{vpnfilter}, focuses on sustaining a direct connection to victim hosts from the C2 server. The attacker employs the notorious VPN Filter malware~\cite{Schneier} to build initial breach on Host 1 and discover Host 2. Then attacker downloads the VPN Filter stage 1 malware from the C2 server to Host 1, subsequently transferring it to Host 2. This malware initiates the download of another executable from the C2 server, executing it to launch the attack and establish a connection with the C2 server. Through this established connection, the attacker transfers a malicious script to Host 2, aimed at gathering sensitive data on the compromised host, as shown in Figure~\ref{fig:3_attacks}(c).

\end{document}